\documentclass[a4paper,onecolumn,11pt,amsmath,amsfonts,unpublished]{quantumarticle}
\pdfoutput=1
\usepackage[utf8]{inputenc}
\usepackage[colorlinks=true,breaklinks=true,allcolors=quantumviolet]{hyperref}

\usepackage[english]{babel}
\usepackage[T1]{fontenc}
\usepackage{amsmath,amssymb,amsthm,amsfonts}
\usepackage{mathtools}
\DeclarePairedDelimiter\abs{\lvert}{\rvert}
\DeclarePairedDelimiter\norm{\lVert}{\rVert}

\DeclarePairedDelimiter\ket{\lvert}{\rangle}
\let\originalleft\left
\let\originalright\right
\renewcommand{\left}{\mathopen{}\mathclose\bgroup\originalleft}
\renewcommand{\right}{\aftergroup\egroup\originalright}

\usepackage{array}
\usepackage{bbm}
\usepackage{bm}
\usepackage{xcolor}

\numberwithin{equation}{section}
\newtheorem{theorem}{Theorem}[section]
\newtheorem{proposition}[theorem]{Proposition}
\newtheorem{lemma}[theorem]{Lemma}
\newtheorem{corollary}[theorem]{Corollary}
\theoremstyle{definition}
\newtheorem{definition}[theorem]{Definition}
\newtheorem{example}[theorem]{Example}
\theoremstyle{remark}
\newtheorem{remark}[theorem]{Remark}

\newcommand{\M}{\mathcal{M}}
\newcommand{\Hs}{\mathcal{H}}
\newcommand{\ip}[2]{\langle #1,#2\rangle}
\newcommand{\Fs}{F^{\mathrm{SLD}}}
\newcommand{\Fb}{F^{\mathrm{BKM}}}
\newcommand{\Tr}{\operatorname{Tr}}
\newcommand{\Var}{\operatorname{Var}}
\newcommand{\Pb}{\mathbb{P}}
\newcommand{\mathe}{\mathrm{e}}
\newcommand{\mathi}{\mathrm{i}}
\newcommand{\total}{\mathop{}\!\mathrm{d}}
\newcommand{\1}{\mathbbm{1}}
\renewcommand{\vec}[1]{\bm{#1}}

\begin{document}

\title{Local Quantum Metrology: Precision Limits for Observers in QFT}

\author{Markus B. Fröb}
\affiliation{Department Mathematik, Friedrich-Alexander-Universit{\"a}t Erlangen-N{\"u}rnberg, Cauerstra{\ss}e 11, 91058 Erlangen, Germany}
\email{markus.froeb@fau.de}
\orcid{0000-0001-6615-9901}

\author{Albert Much}
\affiliation{Quantum Information Research Group, Department of Applied Physics and Astronomy, College of Sciences, University of Sharjah, P. O. Box 27272, Sharjah, United Arab Emirates}
\email{amuch@sharjah.ac.ae}
\orcid{0000-0002-2648-2058}

\maketitle

\begin{abstract}
How precisely can an observer estimate a parameter related to a quantum field? This question is answered for quantum-mechanical systems using the SLD--Bures--Fisher metric and the quantum Cram\'er--Rao bound. However, local algebras in relativistic quantum field theory are of type III: they carry no trace, so that states are not described by density matrices in the algebra, and they have no pure normal states, such that the SLD quantum Fisher information and the quantum Cram\'er--Rao bound cannot be defined in the usual way. We define the SLD (Bures) quantum Fisher information on arbitrary von Neumann algebras through Uhlmann's transition probability and prove a quantum Cram\'er--Rao bound for collective measurements. We also show that the bound is sharp, and equals the largest classical Fisher information obtainable by measurements in the region.

\smallskip

\emph{We dedicate this work to the memory of Armin Uhlmann (1930--2026), a giant on whose shoulders we stand.}
\end{abstract}

\section{Introduction}

Quantum information theory has turned entanglement and the other non-classical features of quantum states from conceptual puzzles into resources for computation, communication and sensing. \emph{Quantum sensing} uses a quantum system as a probe: the quantity to be measured, such as a magnetic field, a frequency or a change in length, is imprinted on the state of the probe and then inferred from measurements on the probe~\cite{DegenReinhardCappellaro2017}. The attainable precision is limited by the quantum fluctuations of the probe state, which make individual measurement outcomes random even when the probe is prepared perfectly. Non-classical probe states, for instance entangled or squeezed states, reduce these fluctuations and reach a precision that no classical probe with the same resources can achieve~\cite{PezzeSmerziOberthalerSchmiedTreutlein2018}. Among the applications of quantum information, sensing is the closest to experimental practice: while useful quantum computation still awaits large-scale error correction~\cite{Preskill2018,Preskill2025}, sensing needs neither error correction nor long-lived quantum memories and already outperforms classical methods in operating instruments~\cite{DegenReinhardCappellaro2017}.

The most prominent application of sensing is the detection of gravitational waves. A passing wave shifts the relative phase of the light in the two arms of a laser interferometer, and the precision of the phase estimate is limited by vacuum fluctuations that enter through the unused input port. Caves proposed to replace this vacuum by a squeezed vacuum, whose reduced fluctuations in the signal quadrature lower the noise below the shot-noise level~\cite{Caves1981}. Squeezed light is now used routinely in gravitational-wave detectors~\cite{AbadieEtAl2011,AasiEtAl2013,TseEtAl2019,GanapathyEtAl2023}. This is a prominent instance in which a quantum resource improves a working scientific instrument, and quantum sensing is expected to have a similar impact on timekeeping, navigation, biomedical imaging and precision tests of fundamental physics~\cite{DegenReinhardCappellaro2017}.

The theoretical basis of quantum sensing is \emph{quantum metrology}. Consider a family of states $\rho_\theta$ depending on a parameter $\theta$. The parameter is inferred by an \emph{estimator} $\hat\theta$, a function of the measurement outcomes that serves as a guess for $\theta$. The estimator is \emph{unbiased} if its mean equals the true value, $\mathbb{E}_\theta[\hat\theta] = \theta$ for all $\theta$, and \emph{locally unbiased} at $\theta_0$ if this holds to first order around $\theta_0$: $\mathbb{E}_{\theta_0}[\hat\theta] = \theta_0$ and $\partial_\theta \mathbb{E}_\theta[\hat\theta] \big\rvert_{\theta = \theta_0} = 1$. Quantum metrology asks how precisely $\theta$ can be estimated from measurements~\cite{Paris2009,GiovannettiLloydMaccone2006,GiovannettiLloydMaccone2011}. Its central result is the quantum Cram\'er--Rao bound of Helstrom~\cite[Chap.~VIII.4]{Helstrom1976} and Holevo~\cite[Chaps.~III.2 and VI.2]{Holevo1982}, in the geometric form of Braunstein and Caves~\cite{BraunsteinCaves1994}: for $\nu$ independent repetitions, every locally unbiased estimator $\hat\theta$ satisfies
\begin{equation}
\Var_\theta(\hat\theta) \geq \frac{1}{\nu\,F_Q(\theta)},\quad F_Q(\theta)=\Tr(\rho_\theta L_\theta^2),\quad \partial_\theta\rho_\theta = \frac{1}{2}(\rho_\theta L_\theta + L_\theta \rho_\theta) ,
\end{equation}
where $F_Q$ is the quantum Fisher information. The last equation defines the symmetric logarithmic derivative (SLD) $L_\theta$, so $F_Q$ is also called the SLD Fisher information, and the bound is attained by measuring $L_\theta$.

Since $F_Q$ is four times the Bures metric, the infinitesimal form of the fidelity between neighbouring states~\cite{BraunsteinCaves1994}, the bound reduces a question about all conceivable measurements to a geometric property of the curve $\theta\mapsto\rho_\theta$. It also quantifies the advantage of entanglement. If $N$ qubits each acquire a phase $\theta$ independently, their Fisher informations add up to $F_Q=N$, and the standard deviation of the estimate scales as $1/\sqrt N$, the standard quantum limit, as expected from the central limit theorem. If the qubits are prepared in an entangled GHZ\footnote{The Greenberger--Horne--Zeilinger (GHZ) state of $N$ qubits is $(\ket{ 0\cdots0 } + \ket{1\cdots1 })/\sqrt2$. If each qubit acquires the relative phase $\theta$ between $\ket{0}$ and $\ket{1}$, the two branches acquire the relative phase $N\theta$.} state, the phases add up coherently to $N\theta$, so that $F_Q=N^2$ and the standard deviation scales as $1/N$, the Heisenberg limit~\cite{GiovannettiLloydMaccone2006,PezzeSmerzi2009}. Here $N$ counts the probes in a single run and $\nu$ the independent repetitions of the run. For uncorrelated probes only the product $\nu N$ matters, since $N$ independent qubits are $N$ repetitions of a one-qubit experiment; entanglement within a run is what separates the two, giving $\Var_\theta(\hat\theta) \geq 1/(\nu N^2)$. This advantage is fragile: if each qubit is subject to independent noise, such as dephasing or photon loss, the scaling returns to $1/\sqrt N$, and entanglement improves the precision only by a constant factor~\cite{DemkowiczDobrzanskiKolodynskiGuta2012}.

Extending this framework to quantum field theory is not only a matter of generality. Every real sensor is local: it couples to the measured quantity in a bounded region of spacetime and during a finite time window, and the quantities it senses, such as electromagnetic fields or gravitational perturbations, are carried by quantum fields. The basic question of quantum metrology for quantum field theory, or \emph{local quantum metrology}, thus becomes: \emph{how well can an observer confined to a spacetime region estimate a parameter encoded in the global state, and what does the restriction to the region cost?} This is no longer an academic question: electro-optic sampling resolves the vacuum fluctuations of the electric field in femtosecond windows~\cite{RiekSeletskiyMoskalenkoSchmidtKrauspeEckartEggertBurkardLeitenstorfer2015}, including their correlations between spacelike separated regions~\cite{BeneaChelmusSettembriniScalariFaist2019,SettembriniLindelHerterBuhmannFaist2022}.

Existing approaches to metrology with quantum fields expand the field in a discrete set of modes and compute the quantum Fisher information of the reduced Gaussian state of one or two modes~\cite{AhmadiBruschiFuentes2014}. This does not answer the question posed above, for two reasons. First, the algebra of finitely many modes is of type I, such that the truncation discards exactly the structure that distinguishes local algebras in quantum field theory. Second, and more importantly, the modes are not localized: the mode operators do not belong to the algebra of any bounded region, and conversely the algebra of a region is not generated by finitely many modes. The Fisher information of a few modes is therefore neither an upper nor a lower bound on the precision attainable by measurements in a region, and its value depends on the choice of modes, which the theory does not single out. For a few sites of a spin chain, the analogous question has been studied in~\cite{AdaniCavazzoniTekluBordoneParis2024}.
 
Two features of quantum field theory obstruct a direct transfer of the finite-dimensional theory. First, under general assumptions the local algebras $\M(O)$ associated to a spacetime region $O$ are type III$_1$ factors~\cite{Haag1996,BuchholzDAntoniFredenhagen1987,Witten2018}: there is no trace, and hence no density matrix affiliated with the algebra\footnote{\label{foot:normal-state}Every normal state of a von Neumann algebra $\M \subset B(\Hs)$ is of the form $x \mapsto \Tr(\rho x)$ for some density matrix $\rho$ on $\Hs$. However, $\rho$ is in general highly non-unique, and for a type III algebra it cannot be chosen in or affiliated with $\M$, so the equation defining the SLD cannot be posed in $\M$. Moreover, a type III factor has no minimal projections and therefore no pure normal states: every normal state of a local algebra is mixed, even if it is the restriction of a vector state such as the vacuum.}; consequently neither the SLD, nor its eigenprojections, nor $\Tr(\rho_\theta L_\theta^2)$ have a meaning within $\M(O)$. Second, by the Reeh--Schlieder theorem~\cite{ReehSchlieder1961,Haag1996} the vacuum, and every vector of bounded energy, is cyclic and separating for the algebra of every region whose causal complement has nonempty interior; such states are entangled across every boundary~\cite{Witten2018}. The complementary region thus acts as an ancilla\footnote{An \emph{ancilla} is an auxiliary quantum system that is not itself the object of interest but is prepared, possibly entangled with the probe, and measured jointly with it.} that is always entangled with the region but inaccessible to the local observer, much like the environment in the noisy metrology discussed above. What replaces the SLD Fisher information, and how much information about $\theta$ is lost because the complement cannot be measured?
 
By a measurement in a region $O$ we mean a positive operator-valued measure (POVM) with effects contained in the local algebra $\M(O)$. We do not model the measurement process itself. Physically realizable local measurements are described by the framework of Fewster and Verch~\cite{FewsterVerch2020}, in which a probe field is coupled to the system inside a compact region; the induced measurements are POVMs in a local algebra, so our bounds apply to them (Remark~\ref{rem:FV}). Conversely, one may ask whether measurements approaching the bound can be realized by probes. The Fisher information depends on the full outcome distribution of a single run, so realizability at the level of expectation values~\cite{FewsterJubbRuep2023} is not sufficient. For the free scalar field, by now a lot is known: full outcome distributions are realized, asymptotically, for Gaussian-modulated measurements of smeared fields~\cite{MandryschNavascues2025}, for every local observable there exists an exact measurement scheme with compact coupling zone~\cite{Fewster2026}, and every POVM with finitely many outcomes is asymptotically realizable, even without disturbing the observables that commute with it~\cite{MandryschSimmonsNavascues2026}. Since our bound is approached by projective measurements with finitely many outcomes (Lemma~\ref{lem:fisher-bound-supremum}), it is approached by Fewster--Verch measurements (Remark~\ref{rem:FVattain}). Independently of any probe model, we show that Gaussian-modulated measurements approach the bound on every von Neumann algebra (Cor.~\ref{cor:gauss}), and that for coherent states of the free scalar field the sharp measurement of a smeared field operator achieves it exactly (Prop.~\ref{prop:fieldopt}).

A Riemannian metric on parametric models of normal positive functionals on $W^*$-algebras, which reduces to the Bures--Helstrom metric for $B(\Hs)$, was constructed from the Jordan product in~\cite{CiagliaDiNoceraJostSchwachhoefer2024}, where a Cram\'er--Rao bound in this setting was left as an open problem. Here we obtain the metric from Uhlmann's transition probability, express it through modular objects, and prove the Cram\'er--Rao bound together with its sharpness.

In order to generalize all these quantities and results to general von Neumann algebras, we start from Uhlmann's transition probability $P(\omega_1,\omega_2)$~\cite{Uhlmann1976,Alberti1983}. It is defined for states on arbitrary $*$-algebras, needs no trace, and for density matrices is equal to the squared fidelity: $\sqrt{P(\rho,\sigma)} = \Tr \abs{ \sqrt{\rho} \sqrt{\sigma} }$. Following the relation between the Bures distance and the SLD Fisher information~\cite{BraunsteinCaves1994}, for every curve $\theta \mapsto \omega_\theta$ of normal states on a von Neumann algebra $\M$ we define the quantum SLD Fisher information as (Def.~\ref{def:sld})
\begin{equation}
\Fs_\M(\theta) \coloneq 8 \lim_{\varepsilon \to 0} \frac{1-\sqrt{P(\omega_\theta,\omega_{\theta+\varepsilon})}}{\varepsilon^2} .
\end{equation}
For faithful density matrices, this reduces exactly to the quantum Fisher information $\Fs_\M(\theta) = F_Q(\theta) = \Tr(\rho_\theta L_\theta^2)$. In a fixed representation, $\sqrt P$ is given by a supremum over the commutant $\M'$ (Lemma~\ref{lem:commutant}), which contains the algebra of the causal complement and equals it under Haag duality, for example for wedges~\cite{BisognanoWichmann1975,BisognanoWichmann1976}. The complementary region is the purifying ancilla.

Our results are as follows: first, we show the operational significance of $\Fs_\M(\theta)$. It is the largest classical Fisher information obtainable from measurements in the region. We prove that the bound $\Var_\theta(\hat\theta) \geq 1/(\nu\,\Fs_\M(\theta))$ holds for collective measurements on $\nu$ copies (Thm.~\ref{thm:qcrb}), and is approached by measuring one local observable on each copy (Thm.~\ref{thm:qcrb_inf}). We also show that whenever a SLD can be defined in the general setting (Def.~\ref{def:sldop}), its measurement achieves the bound (Thm.~\ref{thm:optimal}). We furthermore provide an explicit formula in terms of modular objects:
\begin{equation}
\Fs_\M(\theta) = 4 \norm{ \Pi_\theta \dot\xi_\theta }^2 = 4 \norm{ (1+S_\theta) (1+\Delta_\theta)^{-1} \dot\xi_\theta }^2,
\end{equation}
where $\xi_\theta$ is the implementing vector of the state $\omega_\theta$ in the positive cone, $\dot\xi_\theta$ is its derivative with respect to $\theta$, $\Pi_\theta$ is the projection on the closure of $\{ x \xi_\theta \colon x = x^* \in \M \}$, and $S_\theta = J_\theta \Delta_\theta^{1/2}$ is the Tomita operator. We give an explicit expression for the SLD Fisher information for unitary families generated by a local observable, both bounded (Cor.~\ref{cor:unitary}) and unbounded (Cor.~\ref{cor:unitary-unbounded}). In a two-qubit example, we show that our formulation using modular theory reproduces known results, namely the textbook SLD Fisher information of the reduced state, and the loss equals four times the squared concurrence (Example~\ref{ex:qubits}). We also show that the Bogoliubov--Kubo--Mori (BKM) Fisher information obtained from the Araki--Uhlmann relative entropy~\cite{Araki1975,Uhlmann1977}, recently studied for local states in quantum field theory~\cite{Chatterjee2026}, is never smaller than $\Fs_\M$ (Prop.~\ref{prop:bkm}). Finally, we derive the quantum SLD Fisher information for a free scalar relativistic quantum field in the Rindler wedge (Thm.~\ref{thm:free-field-wedge}), and prove that it can be achieved either as a supremum of the classical Fisher information over all Gaussian-modulated measurements or from a sharp measurement of a certain smeared field operator (Prop.~\ref{prop:fieldopt}). We recover the relation between quantum SLD Fisher information and dynamic susceptibility found previously for Gibbs states \cite{HaukeHeylTagliacozzoZoller2016}, and show that the BKM Fisher information for this case is strictly larger than the SLD one (Prop.~\ref{prop:bkm-wedge}, which proves that the inequality of Prop.~\ref{prop:bkm} is generally strict.) 

Our results thus identify which information-theoretic quantity a local observer can access: relative entropy between local states underlies the derivation of the semiclassical Einstein equations~\cite{DorauMuch2026}, and its second-order term, the BKM Fisher information, is identified with canonical energy in holography~\cite{LashkariVanRaamsdonk2016}. By Prop.~\ref{prop:bkm} this quantity bounds the precision of local measurements, but in general only $\Fs_\M$ is achieved.

The correspondence between the objects known from finite-dimensional quantum me\-tro\-logy and the generalizations to general von Neumann algebras is given in Table~\ref{tab:dictionary}.
\begin{table}[ht]
\centering\small
\renewcommand{\arraystretch}{1.25}
\begin{tabular}{>{\raggedright\arraybackslash}p{4.3cm}>{\raggedright\arraybackslash}p{6.6cm}>{\raggedright\arraybackslash}p{2.8cm}}
\hline
Finite dimensions & Local algebra $\M$ (type III) & Shown in\\
\hline
density matrix $\rho$ & normal state $\omega$ on $\M$ & \\
fidelity $\Tr\abs{ \sqrt{\rho} \sqrt{\sigma} }$ & $\sqrt{P(\omega_1,\omega_2)}$, Uhlmann's transition probability & Lemma~\ref{lem:commutant}\\
purifying ancilla & commutant $\M'$; the complementary region under Haag duality & Lemma~\ref{lem:commutant}\\
$F_Q=\Tr(\rho_\theta L_\theta^2)$ & $\Fs_\M = 4 \norm{ \Pi_\theta\dot\xi_\theta }^2$ & Theorem~\ref{thm:exist}\\
SLD, $\partial_\theta\rho_\theta = \tfrac{1}{2} \{\rho_\theta,L_\theta\}$ & $L_\theta$ affiliated with $\M$, $L_\theta\xi_\theta = 2 \Pi_\theta \dot\xi_\theta$ & Definition~\ref{def:sldop}\\
QCRB $\Var \geq 1/(\nu F_Q)$ & same bound, collective measurements in $\M^{\otimes\nu}$ & Theorem~\ref{thm:qcrb}\\
optimal: eigenprojections of the SLD & spectral measure of an affiliated SLD; bounded local observables come arbitrarily close & Theorem~\ref{thm:optimal}, Lemma~\ref{lem:fisher-bound-supremum} \\
SLD is the smallest monotone metric & $\Fb_\M \geq \Fs_\M$ via the Araki--Uhlmann relative entropy & Proposition~\ref{prop:bkm}\\
Fuchs--Caves: Bures $=$ optimal Hellinger & same, with measurements in $\M$ & Lemma~\ref{lem:infimum} and the remark after it\\
\hline
\end{tabular}
\caption{Finite-dimensional quantum metrology and its counterpart for a local algebra $\M$. We note that a normal state of $\M$ is still given by a density matrix on the ambient Hilbert space, but not by one in or affiliated with $\M$, as explained in footnote~\ref{foot:normal-state}.}
\label{tab:dictionary}
\end{table}

\section{Definition of SLD and Bures--Fisher metric for general von Neumann algebras}

Let $\M$ be a von Neumann algebra acting on a Hilbert space $\Hs$. Given a vector $\xi \in \Hs$, the vector state $\omega_\xi$ realized or implemented by $\xi$ is given by $\omega_\xi(x) = \ip{\xi}{x\xi}$ for $x \in \M$.

\subsection{Transition probability} Uhlmann~\cite[Eq.~(4)]{Uhlmann1976} defines the transition probability $P(\omega_1,\omega_2)$ between two states $\omega_1,\omega_2$ of a $*$-algebra as
\begin{equation}
P(\omega_1,\omega_2) \coloneq \sup \abs{ \ip{\xi_1}{\xi_2} }^2 ,
\end{equation}
where the supremum runs over all representations of the $*$-algebra in which $\omega_1$ and $\omega_2$ are realized as vector states by vectors $\xi_1$ and $\xi_2$, see  also~\cite[Thm.~1]{Alberti1983} and~\cite{UhlmannCrell2009}. The transition probability $P$ satisfies $0 \leq P \leq 1$, is symmetric, and it holds that $P(\omega_1,\omega_2) = 1$ if and only if $\omega_1=\omega_2$. It is furthermore monotone, $P(\omega_1\circ T,\omega_2\circ T) \geq P(\omega_1,\omega_2)$ for unital $2$-positive (in particular completely positive) maps $T \colon \mathcal{N} \to \M$~\cite[Cor.~4]{Alberti1983}, and the squared Bures distance $d^2_\mathrm{B}$ between $\omega_1$ and $\omega_2$ is given by
\begin{equation}
d^2_\mathrm{B}(\omega_1,\omega_2) \coloneq  { 2 - 2 \sqrt{P(\omega_1,\omega_2)} } .
\end{equation}
For explicit computations, we use the following characterization in a fixed representation:
\begin{lemma}[Commutant formula~{\cite[Cor.~1 and 2]{Alberti1983}}]
\label{lem:commutant}
Let $\omega_1, \omega_2$ be realized as vector states for vectors $\xi_1, \xi_2 \in \Hs$. Then it holds that
\begin{equation}
\label{eq:alberti} 
\sqrt{P(\omega_1,\omega_2)} = \sup_{x' \in \M' \colon \norm{x'} \leq 1} \abs{ \ip{\xi_1}{x'\xi_2} } = \sup_{u' \in \M' \colon (u')^* u' = 1} \abs{ \ip{\xi_1}{u'\xi_2} }.
\end{equation} 
\end{lemma}
In particular, every contraction $x'\in\M'$ gives the lower bound $\sqrt{P(\omega_1,\omega_2)} \geq \abs{ \ip{\xi_1}{x'\xi_2} }$. We note that every normal state is a vector state in the standard form of $\M$~\cite{Haagerup1975}, and in any representation in which $\M$ has a separating vector~\cite[Thm.~7.2.3]{KadisonRingrose1986}. In particular, by the Reeh--Schlieder theorem~\cite{ReehSchlieder1961,Haag1996} this applies to the von Neumann algebras $\M(\mathcal{O})$ of local regions $\mathcal{O}$ of Minkowski spacetime in the GNS representation of the Minkowski vacuum state. On the other hand, for $\M = B(\Hs)$ the commutant is given by $\M' = \mathbb{C} 1$, and only pure states are vector states, such that $ P(\omega_\xi,\omega_\eta) = \abs{ \ip{\xi}{\eta} }^2$. Mixed states require an enlarged representation, known as purification, which we compute in the next paragraph.

\subsection{Relation to fidelity and Bures distance in quantum mechanics}

For readers used to quantum mechanics and density matrices, here is the correspondence: Let $\M = B(\Hs)$ with $\dim \Hs = n < \infty$, and let $\rho, \sigma$ be density matrices. Their canonical purifications are vector states on the bipartite Hilbert space $\Hs \otimes \Hs$, which read
\begin{equation}
\label{eq:canonical-purification}
\xi \coloneq (\sqrt\rho\otimes 1) \ket{\Gamma},\qquad \eta \coloneq (\sqrt\sigma\otimes 1)\ket{\Gamma},\qquad \ket{\Gamma} \coloneq \sum_{i=1}^n \ket{i} \otimes \ket{i} ,
\end{equation}
where $\Gamma$ is the (unnormalized) maximally entangled vector with respect to an orthonormal basis $\{ \ket{i} \}_{i=1}^n$ of $\Hs$. Different purifications of the same state differ by a unitary $1\otimes W$ acting on the ancilla, that is, by a unitary of the commutant $\M' = 1 \otimes B(\Hs)$~\cite[Thm.~5.1.1]{Wilde2017}. Therefore, the most general purification of $\rho$ reads $\xi = (\sqrt\rho \otimes U)\ket{\Gamma}$ and the most general purification of $\sigma$ reads $\eta = (\sqrt\sigma \otimes V)\ket{\Gamma}$, where $U$ and $V$ are unitaries. Applying Lemma~\ref{lem:commutant}, we obtain
\begin{equation}
\begin{split}
\sqrt{P(\rho,\sigma)} &= \sup_{x'\in\M' \colon (x')^* x' = 1} \abs{ \ip{\xi}{x'\eta} } \\
&= \sup_{W\in B(\Hs) \colon W^* W = 1} \abs*{ \ip{(\sqrt\rho\otimes U)\ket{\Gamma} }{(1 \otimes W) (\sqrt\sigma\otimes V)\ket{\Gamma} } } \\
&= \sup_{W\in B(\Hs) \colon W^* W = 1} \abs*{ \Tr\left( \sqrt{\rho} \sqrt{\sigma} ( W V U^* )^\top \right) } = \Tr\left( \abs*{ \sqrt{\rho} \sqrt{\sigma} } \right) ,
\end{split}
\end{equation}
since $( W V U^* )^\top$ is again unitary, and the supremum is attained by choosing it to cancel the phase in the polar decomposition of $\sqrt{\rho} \sqrt{\sigma}$. We thus obtain Uhlmann's theorem~\cite{Uhlmann1976}:
\begin{equation}
\label{eq:uhlmann-theorem}
\sqrt{P(\rho,\sigma)} = \max_{(\xi,\eta) \text{ purifications of } (\rho,\sigma)} \abs{ \ip{\xi}{\eta} } = \Tr\abs{ \sqrt{\rho} \sqrt{\sigma} } .
\end{equation}
Thus $\sqrt{P}$ is Uhlmann's fidelity, and $P$ is the largest transition probability that any purifications of the two states can have. Clearly, $\norm{ \xi-\eta }^2 = 2 - 2\,\Re\ip{\xi}{\eta}$ is minimized when $\Re\ip{\xi}{\eta} = \sqrt{P}$, which gives the (squared) Bures distance
\begin{equation}
\label{eq:bures-distance}
d_\mathrm{B}(\rho,\sigma)^2 \coloneq \min_{(\xi,\eta) \text{ purifications of } (\rho,\sigma)} \norm{ \xi-\eta }^2 = 2 - 2 \sqrt{P(\rho,\sigma)} .
\end{equation}
Braunstein and Caves~\cite{BraunsteinCaves1994} showed that the second variation of the Bures distance is the SLD quantum Fisher information. Given a family of faithful density matrices $\rho_\theta$ with $\theta \in \mathbb{R}$, they proved that
\begin{equation}
4 \lim_{\varepsilon \to 0} \frac{d_\mathrm{B}(\rho_\theta,\rho_{\theta+\varepsilon})^2}{\varepsilon^2} = F_Q(\theta) = \Tr(\rho_\theta L_\theta^2),
\end{equation}
where $L_\theta$ is the symmetric logarithmic derivative (SLD) uniquely defined by the relation $\partial_\theta \rho_\theta = \tfrac{1}{2} \left( \rho_\theta L_\theta + L_\theta \rho_\theta \right)$. For the generalization to the non-faithful case, see~\cite{Safranek2016}. We note that for pure states $P = \abs{ \ip{\xi}{\eta} }^2$ and $d_\mathrm{B}$ induces the Fubini--Study metric.

\subsection{Quantum Fisher information for general von Neumann algebras}

As we have seen in the last subsections, Uhlmann's fidelity can be defined for general von Neumann algebras, and reduces to the usual fidelity $\Tr \abs{ \sqrt{\rho} \sqrt{\sigma} }$ on finite-dimensional Hilbert spaces. We can therefore define the quantum Fisher information on a general von Neumann algebra analogously to the finite-dimensional case, replacing the fidelity by Uhlmann's fidelity.

As in quantum estimation theory~\cite{Helstrom1976,Holevo1982,BraunsteinCaves1994}, we consider a curve $\theta\mapsto\omega_\theta$ of states on $\M$.
\begin{definition}
\label{def:sld}
Let $\mathbb{R} \ni \theta\mapsto\omega_\theta$ be a curve of normal states on the von Neumann algebra $\M$. Whenever the limit exists, we define the SLD or Bures quantum Fisher information
\begin{equation}
\Fs_\M(\theta) \coloneq 8 \lim_{\varepsilon \to 0} \frac{1-\sqrt{P(\omega_\theta,\omega_{\theta+\varepsilon})}}{\varepsilon^2} = 4 \lim_{\varepsilon \to 0} \frac{d_\mathrm{B}(\omega_\theta,\omega_{\theta+\varepsilon})^2}{\varepsilon^2}.
\end{equation}
\end{definition}

In general, neither the map $\theta\mapsto\omega_\theta$ nor the transition probability $P(\omega_\theta,\omega_{\theta+\varepsilon})$ need to be differentiable, and thus it is not clear that the limit exists at all. However, we show that $\Fs_\M(\theta)$ is well-defined at least if the family of canonical implementing vectors $\xi_\theta$ of $\omega_\theta$ is differentiable in norm. In the quantum-mechanical setting, these canonical vectors correspond to the canonical purifications~\eqref{eq:canonical-purification}.

\begin{theorem}
\label{thm:exist}
With the notation of Def.~\ref{def:sld}, let $\xi_\theta \in \Hs$ be vectors implementing $\omega_\theta$ which are cyclic and separating for $\M$, and such that the map $\theta \mapsto \xi_\theta$ is continuously differentiable in norm. Then
\begin{equation}
\label{eq:Fexist}
\Fs_\M(\theta) = 4 \norm{ \Pi_\theta \dot\xi_\theta }^2 ,
\end{equation}
where $\dot\xi_\theta \coloneq \partial_\theta \xi_\theta$ and $\Pi_\theta$ is the real-orthogonal projection on the real subspace
\begin{equation}
\label{eq:Htheta-def}
H_\theta \coloneq \overline{\{ x \xi_\theta \colon x = x^* \in \M\}}^{\,\mathbb{R}} \subset \Hs .
\end{equation}
In particular, this applies to the canonical implementing vectors $\xi_\theta \in \mathcal{P}$ of faithful states $\omega_\theta$, whenever these are continuously differentiable in norm.
\end{theorem}
\begin{proof}
Since $\xi_\theta$ is cyclic and separating for $\M$, it is also cyclic and separating for the commutant $\M'$. By Tomita--Takesaki theory~\cite{Longo2008,CorreaDaSilvaLechnerLongo2026}, with respect to the real part of the scalar product we have the decomposition
\begin{equation}
\label{eq:standard-subspace-decomposition}
\Hs = H_\theta \oplus_\mathbb{R} \overline{\{ \mathi x' \xi_\theta \colon x' = (x')^* \in \M'\} }^{\,\mathbb{R}} \eqcolon H_\theta \oplus_\mathbb{R} \mathi H_\theta' .
\end{equation}
That is, every vector $\psi \in \Hs$ can be uniquely written in the form $\psi = \psi_\theta + \psi_\theta'$ with $\psi_\theta = \Pi_\theta \psi$ and $\Re \ip{\psi_\theta}{\psi_\theta'} = 0$, and there are sequences $(x_n)_{n \in \mathbb{N}}$ with $x_n = x_n^* \in \M$ and $(x'_n)_{n \in \mathbb{N}}$ with $x'_n = (x'_n)^* \in \M'$ such that $\lim_{n \to \infty} \norm{ \psi_\theta - x_n \xi_\theta } = \lim_{n \to \infty} \norm{ \psi_\theta' - \mathi x'_n \xi_\theta } = 0$. Both $H_\theta$ and its real-orthogonal complement $\mathi H_\theta'$ are so-called \emph{standard subspaces} of $\Hs$.

Let $u' \coloneq \mathe^{- \mathi \varepsilon x'}$ with $x'=(x')^* \in \M'$ be a unitary from the commutant. By Lemma~\ref{lem:commutant}, Uhlmann's fidelity is a supremum over unitaries in $\M'$, such that we obtain
\begin{equation}
\sqrt{ P(\omega_\theta, \omega_{\theta+\varepsilon}) } \geq \abs{ \ip{\xi_\theta}{u' \xi_{\theta+\varepsilon}} } \geq \Re \ip{\xi_\theta}{u'\xi_{\theta+\varepsilon}} = 1 - \frac{1}{2} \norm{ \xi_\theta - u' \xi_{\theta+\varepsilon} }^2 .
\end{equation}
Since both $u'$ and $\xi_{\theta+\varepsilon}$ are continuously differentiable in norm, it follows that
\begin{equation}
\limsup_{\varepsilon \to 0} \frac{1 - \sqrt{P(\omega_\theta,\omega_{\theta+\varepsilon})}}{\varepsilon^2} \leq \frac{1}{2} \limsup_{\varepsilon \to 0} \frac{\norm{ \xi_\theta-u' \xi_{\theta+\varepsilon} }^2}{\varepsilon^2} = \frac{1}{2} \norm{ - \dot \xi_\theta + \mathi x' \xi_\theta }^2 .
\end{equation}
Decomposing then $\dot \xi_\theta = \Pi_\theta \dot \xi_\theta + (1-\Pi_\theta) \dot \xi_\theta$ with $(1-\Pi_\theta) \dot \xi_\theta \in \mathi H_\theta'$ and choosing a sequence $(x'_n)_{n \in \mathbb{N}}$ with $\lim_{n \to \infty} \norm{ (1-\Pi_\theta) \dot \xi_\theta - \mathi x'_n \xi_\theta } = 0$, we obtain the upper bound
\begin{equation}
\limsup_{\varepsilon \to 0} \frac{1 - \sqrt{P(\omega_\theta,\omega_{\theta+\varepsilon})}}{\varepsilon^2} \leq \frac{1}{2} \norm{ \Pi_\theta \dot \xi_\theta }^2 .
\end{equation}

For the corresponding lower bound, we employ the variational formula~\cite[Cor.~2 (5)]{AlbertiUhlmann2000}
\begin{equation}
\sqrt{ P(\omega_\theta, \omega_{\theta+\varepsilon}) } = \frac{1}{2} \inf_{x \in \M \colon x > 0} \left[ \omega_\theta(x) + \omega_{\theta+\varepsilon}\left( x^{-1} \right) \right] .
\end{equation}
Choosing $x = (1 + \varepsilon y)^2$ with $y = y^* \in \M$, which is positive for all $\abs{\varepsilon} < 1/\norm{ y }$, we obtain
\begin{equation}
\begin{split}
2 \sqrt{ P(\omega_\theta, \omega_{\theta+\varepsilon}) } &\leq \omega_\theta( (1 + \varepsilon y)^2 ) + \omega_{\theta+\varepsilon}\left( (1 + \varepsilon y)^{-2} \right) \\
&= 2 + \varepsilon^2 \norm{ y \xi_\theta }^2 + \norm{ (1 + \varepsilon y)^{-1} \left( \xi_{\theta+\varepsilon} - \xi_\theta \right) }^2 - \norm{ \xi_{\theta+\varepsilon} - \xi_\theta }^2 \\
&\quad+ \left\langle \xi_\theta, \left[ (1 + \varepsilon y)^{-2} - 1 + 2 \varepsilon y \right] \xi_\theta \right\rangle + 2 \Re \left\langle \xi_{\theta+\varepsilon} - \xi_\theta, \left[ (1 + \varepsilon y)^{-2} - 1 \right] \xi_\theta \right\rangle ,
\end{split}
\end{equation}
using that $\norm{ \xi_{\theta+\varepsilon} - \xi_\theta }^2 + 2 \Re \ip{\xi_{\theta+\varepsilon} - \xi_\theta}{\xi_\theta} = \norm{ \xi_{\theta+\varepsilon} }^2 - \norm{ \xi_\theta }^2 = 0$. Since $\xi_{\theta+\varepsilon}$ is continuously differentiable in norm, it follows that
\begin{equation}
\begin{split}
\liminf_{\varepsilon \to 0} \frac{1 - \sqrt{P(\omega_\theta,\omega_{\theta+\varepsilon})}}{\varepsilon^2} &\geq 2 \Re \ip{\dot \xi_\theta}{y \xi_\theta} - 2 \norm{ y \xi_\theta }^2 \\
&= 2 \Re \ip{(1-\Pi_\theta) \dot \xi_\theta}{y \xi_\theta} + \frac{1}{2} \norm{ \Pi_\theta \dot \xi_\theta }^2 - \frac{1}{2} \norm{ \Pi_\theta \dot \xi_\theta - 2 y \xi_\theta }^2 .
\end{split}
\end{equation}
Choosing a sequence $(y_n)_{n \in \mathbb{N}}$ with $\lim_{n \to \infty} \norm{ \Pi_\theta \dot \xi_\theta - 2 y_n \xi_\theta } = 0$, we obtain the bound
\begin{equation}
\liminf_{\varepsilon \to 0} \frac{1 - \sqrt{P(\omega_\theta,\omega_{\theta+\varepsilon})}}{\varepsilon^2} \geq \Re \ip{(1-\Pi_\theta) \dot \xi_\theta}{\Pi_\theta \dot \xi_\theta} + \frac{1}{2} \norm{ \Pi_\theta \dot \xi_\theta }^2 = \frac{1}{2} \norm{ \Pi_\theta \dot \xi_\theta }^2 ,
\end{equation}
and the conclusion.
\end{proof}

\begin{corollary}[Equivalent forms]
\label{cor:equivforms}
With the notation and assumptions of Thm.~\ref{thm:exist}, define for $x = x^* \in \M$
\begin{equation}
\dot\omega_\theta(x) \coloneq \partial_\theta \omega_\theta(x) = 2 \Re \ip{\dot\xi_\theta}{x \xi_\theta}, \quad \Var_{\omega_\theta}(x) \coloneq \omega_\theta(x^2) - \omega_\theta(x)^2 .
\end{equation}
Then it holds that
\begin{equation}
\label{eq:equivforms}
\Fs_\M(\theta) = 4 \inf_{x'= (x')^* \in \M'} \norm{ \dot\xi_\theta - \mathi x' \xi_\theta }^2 = \sup_{x = x^* \in \M \colon \Var_{\omega_\theta}(x)>0} \frac{\dot\omega_\theta(x)^2}{\Var_{\omega_\theta}(x)} .
\end{equation}
\end{corollary}
\begin{proof}
To show the first equality, we use the decomposition~\eqref{eq:standard-subspace-decomposition} of the Hilbert space and write
\begin{equation}
\norm{ \dot\xi_\theta - \mathi x' \xi_\theta }^2 = \norm{ \Pi_\theta \dot\xi_\theta }^2 + \norm{ (1-\Pi_\theta) \dot\xi_\theta - \mathi x' \xi_\theta }^2 + 2 \Re \ip{\Pi_\theta \dot\xi_\theta}{(1-\Pi_\theta) \dot\xi_\theta - \mathi x' \xi_\theta} .
\end{equation}
Since $\mathi x' \xi_\theta \in \mathi H_\theta'$ and $\Pi_\theta$ is orthogonal with respect to the real part of the scalar product, the last term vanishes. Choosing then a sequence $(x'_n)_{n \in \mathbb{N}}$ with $x'_n \in \M'$ such that $\lim_{n \to \infty} \norm{ (1-\Pi_\theta) \dot \xi_\theta - \mathi x'_n \xi_\theta } = 0$, we see that the stated infimum is obtained.

For the second equality, we differentiate $\norm{ \xi_\theta }^2 = 1$ to obtain $\Re \ip{\dot \xi_\theta}{\xi_\theta} = 0$ and write
\begin{equation}
\begin{split}
\sup_{x = x^* \in \M \colon \Var_{\omega_\theta}(x) > 0} \frac{\dot\omega_\theta(x)^2}{\Var_{\omega_\theta}(x)} &= \sup_{x = x^* \in \M \colon \left( x - \omega_\theta(x) \right) \xi_\theta \neq 0} \abs*{ \frac{2 \Re \ip{\dot\xi_\theta}{\left( x - \omega_\theta(x) \right) \xi_\theta}}{\norm{ \left( x - \omega_\theta(x) \right) \xi_\theta }} }^2 \\
&= 4 \sup_{0 \neq h \in H_\theta} \abs*{ \frac{\Re \ip{\dot\xi_\theta}{h}}{\norm{h}} }^2 = 4 \sup_{0 \neq h \in H_\theta} \frac{\bigl( \Re \ip{\Pi_\theta \dot\xi_\theta}{h} \bigr)^2}{\norm{h}^2} \\
&= 4 \norm{ \Pi_\theta \dot\xi_\theta }^2 + 4 \sup_{0 \neq h \in H_\theta} \frac{\bigl( \Re \ip{\Pi_\theta \dot\xi_\theta}{h} \bigr)^2 - \norm{ \Pi_\theta \dot\xi_\theta }^2 \norm{h}^2}{\norm{h}^2} ,
\end{split}
\end{equation}
where we used for the second equality that centering does not decrease the quotient, since $\Re \ip{\dot\xi_\theta}{\left( x - \omega_\theta(x) \right) \xi_\theta} = \Re \ip{\dot\xi_\theta}{x \xi_\theta}$ and $\norm{ \left( x - \omega_\theta(x) \right) \xi_\theta } \leq \norm{ x \xi_\theta }$, and that $H_\theta$ is the closure of the vectors $x \xi_\theta$ with $x = x^* \in \M$. For the third equality we used that $\Pi_\theta$ is orthogonal with respect to the real part of the scalar product. By the Cauchy--Schwarz inequality, we have $\bigl( \Re \ip{\Pi_\theta \dot\xi_\theta}{h} \bigr)^2 \leq \norm{ \Pi_\theta \dot\xi_\theta }^2 \norm{h}^2$, with equality for $h = \Pi_\theta \dot\xi_\theta$, and we see that also the stated supremum is obtained.
\end{proof}

The second form is particularly instructive: it shows that the quantum Fisher information is the best signal-to-noise ratio that can be obtained by measurements inside the region.

To recover the quantum-mechanical result, we take $\M = B(\Hs) \otimes 1$ acting on the bipartite Hilbert space $\Hs \otimes \Hs$. For a differentiable family of density matrices $\rho_\theta$ of full rank, the canonical implementing vectors are the purifications $\xi_\theta = (\sqrt{\rho_\theta}\otimes 1) \ket{\Gamma}$ \eqref{eq:canonical-purification}. Let $L_\theta = L_\theta^*$ be the symmetric logarithmic derivative, defined by $\dot\rho_\theta = \frac{1}{2} (\rho_\theta L_\theta + L_\theta \rho_\theta)$. For $x = x^* \in B(\Hs)$, it then follows that
\begin{equation}
\begin{split}
\dot \omega_\theta(x \otimes 1) &= 2 \Re \ip{ \dot\xi_\theta }{ (x \otimes 1) \xi_\theta } = 2 \Re \ip{ (\partial_\theta \sqrt{\rho_\theta}\otimes 1) \Gamma }{ (x \otimes 1) (\sqrt{\rho_\theta}\otimes 1) \Gamma } \\
&= 2 \Re \Tr\left( \partial_\theta \sqrt{\rho_\theta} \, x \sqrt{\rho_\theta} \right) = \Re \Tr\left( x \dot \rho_\theta \right) = \frac{1}{2} \Re \Tr\left( x \rho_\theta L_\theta + x L_\theta \rho_\theta \right) \\
&= \Re \Tr\left( x L_\theta \rho_\theta \right) = \Re \ip{(L_\theta\otimes 1)\xi_\theta} {(x\otimes 1)\xi_\theta} .
\end{split}
\end{equation}
Hence we have $\Re \ip{ 2 \dot\xi_\theta - (L_\theta\otimes 1)\xi_\theta }{ (x \otimes 1) \xi_\theta } = 0$, which shows that $2 \dot\xi_\theta - (L_\theta\otimes 1)\xi_\theta$ is real-orthogonal to the standard subspace $H_\theta = \overline{\{ (x\otimes 1) \xi_\theta \colon x = x^* \in B(\Hs) \}}^{\,\mathbb{R}}$. Consequently, it holds that
\begin{equation}
\label{eq:SLD-qm-projection}
\Pi_\theta \dot\xi_\theta = \frac{1}{2} (L_\theta\otimes 1)\xi_\theta ,
\end{equation}
and \eqref{eq:Fexist} becomes
\begin{equation}
\Fs_\M(\theta) = 4 \norm{ \Pi_\theta \dot\xi_\theta }^2 = \norm{ (L_\theta \otimes 1) \xi_\theta }^2 = \Tr(\rho_\theta L_\theta^2) = F_Q(\theta),
\end{equation}
recovering the usual SLD quantum Fisher information.

Let us give yet another expression for the quantum Fisher information, which will be useful for concrete computations.
\begin{theorem}[Modular form]
\label{thm:modform}
With the notation and assumptions of Thm.~\ref{thm:exist}, let $S_\theta = J_\theta \Delta_\theta^{1/2}$ be the Tomita operator of $(\M,\xi_\theta)$. Then the Tomita operator of the standard subspace $H_\theta$ is $S_\theta$, the real orthogonal projection is given by
\begin{equation}
\label{eq:PiThetaModular}
\Pi_\theta = (1+\Delta_\theta)^{-1}+J_\theta\Delta_\theta^{1/2}(1+\Delta_\theta)^{-1} = (1+S_\theta)(1+\Delta_\theta)^{-1},
\end{equation}
and it holds that
\begin{equation}
\label{eq:Fmod}
\begin{split}
\Fs_\M(\theta) &= 4 \norm*{ (1+S_\theta) (1+\Delta_\theta)^{-1} \dot\xi_\theta }^2 \\
&= 4 \ip{\dot\xi_\theta}{(1+\Delta_\theta)^{-1}\dot\xi_\theta} + 4 \Re \ip{\dot\xi_\theta}{S_\theta (1+\Delta_\theta)^{-1}\dot\xi_\theta}.
\end{split}
\end{equation}
\end{theorem}
\begin{proof}
That $S_\theta$ is the Tomita operator of $H_\theta$ follows again from Tomita--Takesaki theory~\cite{Longo2008,CorreaDaSilvaLechnerLongo2026}. The formula for the projection is due to Rieffel and van Daele~\cite[Def.~2.1, Prop.~2.2, App.]{RieffelVanDaele1977}, and also shown in \cite[Prop.~1.3.17]{CorreaDaSilvaLechnerLongo2026}. The relation \eqref{eq:Fmod} then follows from $\norm{ \Pi_\theta v }^2 = \Re \ip{\Pi_\theta v}{\Pi_\theta v} = \Re \ip{v}{\Pi_\theta v}$ and the fact that $\Delta_\theta$ is self-adjoint.
\end{proof}

With this, we can give a large class of examples where the quantum Fisher information can be computed explicitly.
\begin{corollary}[Unitary families with bounded generator]
\label{cor:unitary}
Let $\Omega \in \Hs$ be cyclic and separating for $\M$, let $S = J \Delta^{1/2}$ be the corresponding Tomita operator with the modular conjugation $J$ and the modular operator $\Delta$, let $G = G^* \in \M$, and let $\omega_\theta$ be the state on $\M$ defined by $\omega_\theta(x) \coloneq \left\langle u_\theta \Omega, x \, u_\theta \Omega \right\rangle$ with $u_\theta \coloneq \mathe^{\mathi \theta G}$. Then the assumptions of Thm.~\ref{thm:exist} are fulfilled, and the quantum Fisher information is independent of $\theta$ and given by
\begin{equation}
\Fs_\M = 4 \norm*{ \frac{1-\Delta}{1+\Delta} \,G \Omega }^2 = 4 \norm{ \Pi_0 \mathi G \Omega }^2 ,
\end{equation}
where $\Pi_0$ is the real-orthogonal projection on $H_0 = \overline{\{ x \Omega \colon x = x^* \in \M\}}^{\,\mathbb{R}}$~\eqref{eq:Htheta-def}.
\end{corollary}
\begin{proof}
First note that since $J$ maps $\M$ to the commutant $\M'$, $u'_\theta \coloneq J u_\theta J \in \M'$ is a unitary in the commutant and conjugation with it gives the same state on $\M$:
\begin{equation}
\left\langle u'_\theta u_\theta \Omega, x \, u'_\theta u_\theta \Omega \right\rangle = \left\langle u_\theta \Omega, (u'_\theta)^* x \, u'_\theta u_\theta \Omega \right\rangle = \left\langle u_\theta \Omega, x u_\theta \Omega \right\rangle = \omega_\theta(x)
\end{equation}
for all $x \in \M$. However, $\xi_\theta \coloneq u_\theta u'_\theta \Omega \in \mathcal{P}$ lies in the positive cone~\cite[Thm.~C.1]{ArakiMasuda1982}, and is thus the unique canonical implementing vector. Since $u_\theta$ is a unitary in $\M$, $\xi_\theta$ is cyclic and separating, and we obtain
\begin{equation}
\begin{split}
\dot \xi_\theta &= \mathe^{\mathi \theta G} \, \mathi G \, J \mathe^{\mathi \theta G} J \Omega + \mathe^{\mathi \theta G} J \, \mathi G \mathe^{\mathi \theta G} J \Omega = u_\theta \mathi G u'_\theta \Omega + u_\theta J \, \mathi G J u'_\theta \Omega \\
&= \mathi u_\theta (G - J G J) u'_\theta \Omega = \mathi u_\theta u'_\theta (G - J G J) \Omega ,
\end{split}
\end{equation}
using that the modular conjugation $J$ is antiunitary and satisfies $J^2 = 1$.

Moreover, we compute
\begin{equation}
( S_\theta - u_\theta u'_\theta S u_\theta^* (u'_\theta)^* ) x u_\theta u'_\theta \Omega = S_\theta x u_\theta u'_\theta \Omega - x^* u_\theta u'_\theta \Omega = S_\theta x \xi_\theta - x^* \xi_\theta = 0 ,
\end{equation}
and since $\xi_\theta$ is separating, we obtain $S_\theta = u_\theta u'_\theta S u_\theta^* (u'_\theta)^*$ and from this $\Delta_\theta = S_\theta^* S_\theta = u'_\theta u_\theta \Delta u_\theta^* (u'_\theta)^*$. Using Thm.~\ref{thm:modform} and the fact that $S$ is antilinear, it follows that
\begin{equation}
\begin{split}
\Fs_\M(\theta) &= 4 \norm*{ (1+S_\theta) (1+\Delta_\theta)^{-1} \dot\xi_\theta }^2 = 4 \norm*{u_\theta u'_\theta ( 1 + S ) ( 1 + \Delta )^{-1} u_\theta^* (u'_\theta)^* \dot\xi_\theta }^2 \\
&= 4 \norm*{ ( 1 - S ) ( 1 + \Delta )^{-1} (G - J G J) \Omega }^2 .
\end{split}
\end{equation}
Since $G = G^*$ and $J \Omega = \Omega$, we obtain
\begin{equation}
J G J \Omega = J G \Omega = J S G \Omega = \Delta^{1/2} G \Omega ,
\end{equation}
and using that $J f(\Delta) J = f^*(\Delta^{-1})$ it follows furthermore that
\begin{equation}
S f(\Delta) G \Omega = J \Delta^{1/2} f(\Delta) J J G J \Omega = \Delta^{-1/2} f^*(\Delta^{-1}) J G J \Omega = f^*(\Delta^{-1}) G \Omega .
\end{equation}
Hence we obtain as required
\begin{equation}
\Fs_\M(\theta) = 4 \norm*{ ( 1 - S ) ( 1 + \Delta )^{-1} (1 - \Delta^{1/2}) G \Omega }^2 = 4 \norm*{ \frac{1-\Delta}{1+\Delta} \, G \Omega }^2 .
\end{equation}
On the other hand, we compute using Thm.~\ref{thm:modform} and the above results
\begin{equation}
\norm{ \Pi_0 \mathi G \Omega }^2 = \norm{ (1+S) (1+\Delta)^{-1} \mathi G \Omega }^2 = \norm{ (1-S) (1+\Delta)^{-1} G \Omega }^2 = \norm*{ \frac{1-\Delta}{1+\Delta} G \Omega }^2 ,
\end{equation}
and obtain also the second equality.
\end{proof}

Again it is straightforward to see that we can recover known results by specializing the above to a finite-dimensional system.
\begin{example}[Two qubits]
\label{ex:qubits}
Consider the bipartite Hilbert space $\Hs = \Hs_1 \otimes \Hs_2$ with $\Hs_i = \mathbb{C}^2$, the algebra $\M = B(\Hs_1) \otimes \1$ and the cyclic and separating vector $\Omega = \sqrt{p} \, \ket{00} + \sqrt{1-p} \, \ket{11}$ with $p \in (0,1)$. Let $\xi_\theta = \mathe^{\mathi \theta G} \Omega$ with $G = \sigma_x \otimes \1$, where $\sigma_x = \left( \begin{smallmatrix} 0 & 1 \\ 1 & 0 \end{smallmatrix} \right)$ is the first Pauli matrix, i.e.,
\begin{equation}
\begin{split}
\xi_\theta &= \left( \mathe^{\mathi \theta \sigma_x} \otimes 1 \right) \Omega = \cos \theta \, \Omega + \mathi \sin \theta \, G \Omega \\
&= \cos \theta \left( \sqrt{p}\, \ket{00} + \sqrt{1-p}\, \ket{11} \right) + \mathi \sin \theta \left( \sqrt{p}\, \ket{10} + \sqrt{1-p}\, \ket{01} \right) .
\end{split}
\end{equation}
The local observer then sees the state described by the reduced density matrix
\begin{equation}
\rho_1(\theta) \coloneq \Tr_2\left( \xi_\theta \xi_\theta^\dagger \right) = \frac{1}{2} \1 + \left( p - \frac{1}{2} \right) \sigma_\theta , \quad \sigma_\theta \coloneq \cos(2\theta) \sigma_z + \sin(2\theta) \sigma_y
\end{equation}
with the other Pauli matrices $\sigma_y = \left( \begin{smallmatrix} 0 & - \mathi \\ \mathi & 0 \end{smallmatrix} \right)$ and $\sigma_z = \left( \begin{smallmatrix} 1 & 0 \\ 0 & -1 \end{smallmatrix} \right)$, and the complementary reduced density matrix reads
\begin{equation}
\rho_2 \coloneq \Tr_1\left( \xi_\theta \xi_\theta^\dagger \right) = \begin{pmatrix} p & 0 \\ 0 & 1-p \end{pmatrix} = \frac{1}{2} \1 + \left( p - \frac{1}{2} \right) \sigma_z .
\end{equation}
The modular operator is given by
\begin{equation}
\Delta_\theta = \rho_1(\theta) \otimes \rho_2^{-1} =  \frac{1}{4 p (1-p)} \left[ \mathbbm{1} + (2p-1) \sigma_\theta \right] \otimes \left[ \1 - (2p-1) \sigma_z \right] ,
\end{equation}
and we compute
\begin{equation}
1 - \Delta_\theta - \left( p - \frac{1}{2} \right) \left( \1 \otimes \sigma_z - \sigma_\theta \otimes \1 \right) (1 + \Delta_\theta) = \frac{(2p-1)^2}{8 p (1-p)} ( \sigma_\theta^2 - \1 ) \otimes \left[ \1 - (2p-1) \sigma_z \right] = 0 ,
\end{equation}
using that $\sigma_\theta^2 = \1$. Hence $(1-\Delta_\theta)/(1+\Delta_\theta) = (p-1/2) \left( \1 \otimes \sigma_z - \sigma_\theta \otimes \1 \right)$, and using Cor.~\ref{cor:unitary} it follows that
\begin{equation}
\begin{split}
\Fs_\M &= 4 \norm*{ \frac{1-\Delta}{1+\Delta}\,G\Omega }^2 = (2p-1)^2 \norm*{ \left( \1 \otimes \sigma_z - \sigma_0 \otimes \1 \right) G \Omega }^2 \\
&= (2p-1)^2 \norm*{ \left( \1 \otimes \sigma_z - \sigma_z \otimes \1 \right) ( \sigma_x \otimes \1 ) \Omega }^2 = (2p-1)^2 \norm*{ 2 \sqrt{p} \, \ket{10} - 2 \sqrt{1-p} \, \ket{01} }^2 \\
&= (2p-1)^2 \left[ 4 p + 4 (1-p) \right] = 4 (2p-1)^2 ,
\end{split}
\end{equation}
which agrees with the standard spectral expression for the SLD quantum Fisher information~\cite[Sec.~4.1]{Paris2009} for the family $\rho_1(\theta)$. With respect to the global Fisher information $\Fs_\mathrm{global} = 4 \norm{G\Omega}^2  = 4 \norm{ \Omega}^2 = 4$, the information loss is given by
\begin{equation}
\Fs_\mathrm{global} - \Fs_\M = 4 - 4 (2p-1)^2 = 4 C^2 ,
\end{equation}
where $C = 2 \sqrt{p (1-p)}$ is the concurrence~\cite{Wootters1998}.
\end{example}

We may furthermore extend this class to include unbounded generators.
\begin{corollary}[Unitary families with unbounded generator]
\label{cor:unitary-unbounded}
Let $\Omega \in \Hs$ be cyclic and separating for $\M$, let $\Delta$ be the corresponding modular operator, let $G = G^*$ be affiliated\footnote{A self-adjoint operator $G$ is \emph{affiliated} with $\M$ if all its spectral projections lie in $\M$, equivalently if $\mathe^{\mathi \theta G} \in \M$ for all $\theta \in \mathbb{R}$. } with $\M$ with $\Omega \in D(G)$, and let $\xi_\theta \coloneq u_\theta \Omega$ with $u_\theta \coloneq \mathe^{\mathi \theta G}$. Then the quantum Fisher information of the vector state $\omega_\theta$ with implementing vector $\xi_\theta$ exists, is independent of $\theta$ and given by
\begin{equation}
\Fs_\M = 4 \norm*{ \frac{1-\Delta}{1+\Delta} \,G \Omega }^2 = 4 \norm{ \Pi_0 \mathi G \Omega }^2 ,
\end{equation}
where $\Pi_0$ is the real-orthogonal projection on $H_0 = \overline{\{ x \Omega \colon x = x^* \in \M\}}^{\,\mathbb{R}}$~\eqref{eq:Htheta-def}.
\end{corollary}
\begin{proof}
Since $u_\theta \in \M$ is unitary, $\xi_\theta = u_\theta \Omega$ is cyclic and separating for $\M$, and since $\Omega \in D(G)$, by Stone's theorem the map $\theta \mapsto \xi_\theta$ is continuously differentiable in norm, with $\dot\xi_\theta = \mathi u_\theta G \Omega$. Moreover, $x \xi_\theta = u_\theta ( u_\theta^* x u_\theta ) \Omega$ shows that $H_\theta = u_\theta H_0$ and hence $\Pi_\theta = u_\theta \Pi_0 u_\theta^*$. Thm.~\ref{thm:exist} thus gives
\begin{equation}
\Fs_\M(\theta) = 4 \norm{ u_\theta \Pi_0 u_\theta^* \, \mathi u_\theta G \Omega }^2 = 4 \norm{ \Pi_0 \mathi G \Omega }^2 .
\end{equation}
The spectral truncations $G_n \coloneq E^G([-n,n]) G \in \M$ converge strongly $\lim_{n \to \infty} \norm{ (G_n-G) \Omega } = 0$ and satisfy $S G_n \Omega = G_n \Omega$. Hence, we have $G \Omega \in H_0$ with $S G \Omega = G \Omega$. For bounded Borel functions $f$, on $D(S)$ it holds that $S f(\Delta) = \overline{f}(\Delta^{-1}) S$, and it follows that $S (1+\Delta)^{-1} G \Omega = \Delta (1+\Delta)^{-1} G \Omega$. Using Thm.~\ref{thm:modform} and the antilinearity of $S$, we obtain
\begin{equation}
\Pi_0 \mathi G \Omega = (1+S) (1+\Delta)^{-1} \mathi G \Omega = \mathi (1-S) (1+\Delta)^{-1} G \Omega = \mathi \, \frac{1-\Delta}{1+\Delta} \, G \Omega ,
\end{equation}
which gives both expressions.
\end{proof}

\section{Quantum Cram\'er--Rao bound}
\label{sec:qcrb}

In quantum metrology, the SLD Fisher information bounds the attainable precision through the quantum Cram\'er--Rao bound~\cite{Helstrom1976,Holevo1982,Paris2009}. The bound and its attainability are routinely analysed for qubit and Gaussian probes, see e.g.~\cite{TekluBinaParis2022,AsjadTekluParis2023}. We first recall the finite-dimensional argument of Braunstein and Caves~\cite{BraunsteinCaves1994} and then explain how it carries over to general von Neumann algebras.

\subsection{The argument of Braunstein and Caves}

Let $\theta \mapsto \rho_\theta \in B(\Hs)$ be a curve of density matrices and $E \colon \Sigma \to B(\Hs)$ be a measurement (POVM) with outcome distribution $p_\theta(s) = \Tr( E(s) \rho_\theta )$ for $s \in \Sigma$, the space of possible outcomes. The argument of~\cite{BraunsteinCaves1994} consists of two optimizations. The first is classical: an estimate is computed from the outcomes of $\nu$ repetitions, so its variance obeys the classical Cram\'er--Rao bound $\Var_\theta(\hat\theta) \geq 1/(\nu F^\mathrm{cl}(\theta;E))$, which maximum-likelihood estimation attains asymptotically. The classical Fisher information $F^\mathrm{cl}(\theta;E)$ is therefore the natural measure of how well the outcome distributions distinguish neighbouring values of $\theta$. The second optimization is quantum: $F^\mathrm{cl}(\theta;E)$ is maximized over all possible outcome distributions $p_\theta$, hence all possible measurements $E$. A Cauchy--Schwarz estimate gives $F^\mathrm{cl}(\theta;E) \leq \Tr(\rho_\theta L_\theta^2)$~\cite[Eq.~(24)]{BraunsteinCaves1994} where $L_\theta$ is the symmetric logarithmic derivative defined through  $\partial_\theta \rho_\theta = \tfrac{1}{2} (\rho_\theta L_\theta + L_\theta \rho_\theta)$, with equality if $E$ consists of the eigenprojections of $L_\theta$~\cite[Eqs.~(25)--(27)]{BraunsteinCaves1994}. Hence
\begin{equation}
\max_E F^\mathrm{cl}(\theta;E) = \Tr(\rho_\theta L_\theta^2),
\end{equation}
and Braunstein and Caves identify the resulting metric $\Tr(\rho_\theta L_\theta^2) \total \theta^2$ with four times the Bures metric~\cite[Eqs.~(28)--(31)]{BraunsteinCaves1994}.

On a general von Neumann algebra, in particular one of type III, none of the ingredients of~\cite{BraunsteinCaves1994} is available: there is no trace, so states are not given by density matrices in or affiliated with the algebra, and hence neither the SLD $L_\theta$ nor its eigenprojections can be defined as above. Two objects do survive. The first is the transition probability $P$, which by Lemma~\ref{lem:commutant} is a supremum over the commutant and needs no trace. The second is the outcome distribution of a measurement, which is an ordinary probability measure. We therefore reverse the order of the argument. Braunstein and Caves start from the classical Fisher information, maximize it over measurements and identify the maximum with the Bures metric. We start from the Bures side, i.e., from Def.~\ref{def:sld}, show that it bounds the classical Fisher information of every measurement, and then identify the measurements that reach the bound.

More concretely, we note that a link between the classical and the quantum side is that both Fisher informations are second-order coefficients of a fidelity:
\begin{equation}
B(\Pb_\theta,\Pb_{\theta+\varepsilon}) = 1 - \frac{1}{8} F^\mathrm{cl}(\theta;E) \, \varepsilon^2 + o(\varepsilon^2), \qquad \sqrt{P(\omega_\theta,\omega_{\theta+\varepsilon})} = 1 - \frac{1}{8} \Fs_\M(\theta) \, \varepsilon^2 + o(\varepsilon^2).
\end{equation}
Here $B(\Pb_\theta,\Pb_{\theta+\varepsilon})$ is the Bhattacharyya coefficient of the outcome distribution $\Pb_\theta$ of $E$ defined in \eqref{eq:bhatcoef}, and the second expansion is Def.~\ref{def:sld}. An inequality between the two fidelities at finite $\varepsilon$ therefore becomes an inequality between the two Fisher informations, and no derivative of a density matrix is ever needed. We will see in the next subsections that for a measurement $E$ on $\nu$ copies, the fidelities can be ordered according to
\begin{equation}
\underbrace{ \int \sqrt{p_\theta\,p_{\theta+\varepsilon}} \total \mu }_{1 - \frac{1}{8} F^\mathrm{cl}(\theta;E) \, \varepsilon^2 + o(\varepsilon^2)} \ \geq\ \sqrt{ P\bigl(\omega_\theta^{\otimes\nu},\omega_{\theta+\varepsilon}^{\otimes\nu}\bigr)} = \underbrace{ \sqrt{P(\omega_\theta,\omega_{\theta+\varepsilon})}^{\,\nu} }_{1 - \frac{1}{8} \nu \, \Fs_\M(\theta) \, \varepsilon^2 + o(\varepsilon^2)} .
\end{equation}
Each relation has a simple meaning: The inequality (Lemma~\ref{lem:mono}) is monotonicity under measurements. A measurement maps the states on $\M^{\otimes\nu}$ to classical probability distributions, and in doing so it can only lose information: the outcome distributions are at least as close to each other as the states that produced them. Moreover, we will see that taking the infimum over all measurements yields equality (Lemma~\ref{lem:infimum}).

The equality (Lemma~\ref{lem:product}) concerns $\nu$ independently prepared copies. Their fidelity is at least the $\nu$-th power of the single-copy fidelity, so $\nu$ uncorrelated repetitions cannot be more distinguishable than the product of the individual runs, even when they are measured collectively. However, it turns out that they are exactly as distinguishable since Uhlmann's fidelity is multiplicative over tensor products. Comparing the coefficients of $\varepsilon^2$ at the two ends gives $F^\mathrm{cl}(\theta;E) \leq \nu\,\Fs_\M(\theta)$, and the classical Cram\'er--Rao bound turns this into $\Var_\theta(\hat\theta) \geq 1/(\nu\,\Fs_\M(\theta))$ (Thm.~\ref{thm:qcrb}).

\subsection{Measurements and fidelities}

\begin{definition}[Measurement, {\cite[\S3]{DaviesLewis1970}; cf.~\cite[Def.~26]{HeinosaariZiman2008}}]
\label{def:povm}
A \emph{measurement} (positive operator-valued measure, POVM) in $\M$ with outcome space $(X,\Sigma)$ is a map $E \colon \Sigma \to \M$ assigning an operator $E(s) \in \M$ to each element $s$ of the Borel $\sigma$-algebra $\Sigma$ over $X$, fulfilling the following properties:
\begin{itemize}
\item \emph{(Positivity)} $E(s) \in [0,1]_\M$ for all $s \in \Sigma$, i.e., $E(s)$ is an \emph{effect},
\item \emph{(Almost sure event)} $E(X) = 1$ and $E(\emptyset) = 0$,
\item \emph{(Additivity)} $E(\bigcup_i s_i) = \sum_i E(s_i)$ in the $\sigma$-weak topology for every sequence of pairwise disjoint $s_i \in \Sigma$.
\end{itemize}
For a normal state $\omega$ on $\M$, the \emph{outcome distribution} of $E$ is the probability measure $s \mapsto \omega(E(s))$ on $(X,\Sigma)$. A \emph{projective measurement} is a measurement for which $E(s)$ is a projection for all $s \in \Sigma$. The set of all measurements in $\M$ is denoted by $\mathcal{E}_\M$.
\end{definition} 

\begin{remark}[Relation to the Fewster--Verch framework]
\label{rem:FV}
Quantum metrology does not model the measurement process; it only uses that a measurement is a POVM, whose effects give the outcome probabilities. Fewster and Verch~\cite{FewsterVerch2020} provide a physical description of local measurements that yields such POVMs: A probe, prepared in a state $\sigma$ independent of the system, is coupled to the system in a compact region $K$. A measured probe effect $Q$ induces a system effect $\mathcal{E}_\sigma(Q)$, whose expectation value in the system state is the probability of the probe outcome. The map $\mathcal{E}_\sigma$ is completely positive and unital, and $\mathcal{E}_\sigma(Q)$ is an element of the local von Neumann algebra $\M(L)$ for every connected, open, causally convex region $L\supset K$~\cite{FewsterVerch2020}. A probe measurement with finitely many outcomes therefore induces a POVM in $\M(L)$, and by Thm.~\ref{thm:qcrb} its classical Fisher information is at most $\Fs_{\M(L)}(\theta)$. In particular, the probe acts as an ancilla prepared independently of the system, and cannot add information about $\theta$. We note that only the outcome statistics enter here, but not the state after the measurement. Whether the bound can be approached by such measurements is discussed in Remark~\ref{rem:FVattain}.
\end{remark}
For normal states $\omega_1, \omega_2$, the outcome distributions $\Pb_i \coloneq \omega_i \circ E$ are probability measures on $(X,\Sigma)$. Let $p_i$ be their Radon--Nikod\'ym derivative (density) with respect to a dominating $\sigma$-finite measure $\mu$, for example $\mu = (\Pb_1 + \Pb_2)/2$. The Bhattacharyya coefficient~\cite{Bhattacharyya1943}
\begin{equation}
\label{eq:bhatcoef}
B(\Pb_1,\Pb_2) \coloneq \int_X \sqrt{ p_1 \, p_2 } \total \mu = \left\langle \sqrt{ p_1 }, \sqrt{ p_2 } \right\rangle_{L^2(X,\mu)} = 1 - \frac{1}{2} \norm{ \sqrt{p_1} - \sqrt{p_2} }^2_{L^2(X,\mu)}
\end{equation}
does not depend on $\mu$, and the last equality follows since $\sqrt{ p_i }$ is a unit vector in $L^2(X,\mu)$. Hence, the Bhattacharyya coefficient satisfies $B(\Pb_1,\Pb_2) \in [0,1]$, and it is equal to 1 if and only if $p_1$ and $p_2$ are equal $\mu$-almost everywhere; in particular, we have $B(\Pb_1,\Pb_1) = 1$. It is the classical version of Uhlmann's fidelity $\sqrt{P(\omega_1,\omega_2)}$: for commuting density matrices $\rho,\sigma$ with eigenvalues $\{ \rho_i \}_{i=1}^n$ and $\{ \sigma_i \}_{i=1}^n$, we have \eqref{eq:uhlmann-theorem}
\begin{equation}
\sqrt{ P(\rho,\sigma) } = \Tr\abs{ \sqrt{\rho} \sqrt{\sigma} } = \sum_{i=1}^n \sqrt{ \rho_i \sigma_i } = B(\Pb_\rho,\Pb_\sigma) ,
\end{equation}
where $\Pb_\rho$ is the probability measure assigning the eigenvalue $\rho_i$ to $i \in \{1,\ldots,n\}$, and $\mu$ can be taken as the uniform probability measure on $\{1,\ldots,n\}$. The (squared) Bures distance $d_\mathrm{B}(\rho,\sigma)^2 = 2 - 2 \sqrt{P(\rho,\sigma)}$ \eqref{eq:bures-distance} is then equal to the (squared) Hellinger distance $\norm{ \sqrt{ p_\rho } - \sqrt{ p_\sigma } }_{L^2(X,\mu)}^2 = 2 - 2 B(\Pb_\rho,\Pb_\sigma)$. The quantity $- \ln B$ is known as Bhattacharyya distance~\cite{Kailath1967}, and the angle $\arccos B$ is Wootters' statistical distance~\cite{Wootters1981}, which Braunstein and Caves~\cite{BraunsteinCaves1994} take as their starting point.

We first show that a measurement can only increase the fidelity.
\begin{lemma}[Monotonicity under measurements]
\label{lem:mono}
Let $E \in \mathcal{E}_\M$ be a measurement with outcome space $(X,\Sigma)$, let $\omega_1,\omega_2$ be normal states on $\M$, and let $p_1,p_2$ be the densities of the outcome distributions $\Pb_j = \omega_j \circ E$ with respect to a dominating $\sigma$-finite measure $\mu$ on the outcome space. Then it holds that
\begin{equation}
\label{eq:monotonity-ineq1}
\sqrt{P(\omega_1,\omega_2)} \leq B(\Pb_1,\Pb_2) = \int_X \sqrt{p_1\,p_2} \total \mu .
\end{equation}
\end{lemma}
\begin{proof}
We first note that the right-hand side can be written as an infimum over finite partitions of $X$. Namely, for any partition $X = \bigcup_i X_i$ of $X$ into disjoint sets $X_i$ the Cauchy--Schwarz inequality gives
\begin{equation}
\int_X \sqrt{p_1\,p_2} \total \mu = \sum_i \int_{X_i} \sqrt{p_1\,p_2} \total \mu \leq \sum_i \sqrt{ \int_{X_i} p_1 \total \mu } \, \sqrt{ \int_{X_i} p_2 \total \mu } = \sum_i \sqrt{ \Pb_1(X_i) \, \Pb_2(X_i) } \, .
\end{equation}
We would like to show that
\begin{equation}
\int_X \sqrt{p_1\,p_2} \total \mu = \inf_{X = \bigcup_i X_i, X_i \cap X_j = \emptyset} \sum_i \sqrt{ \Pb_1(X_i) \, \Pb_2(X_i) } \, ,
\end{equation}
i.e., the sum over partitions approaches the integral as it becomes finer. For this, we have to bound the difference between the integral and the right-hand side. By definition of the Lebesgue integral, there exists a sequence of approximations $( p_1^{(n)} )_{n \in \mathbb{N}}$, $( p_2^{(n)} )_{n \in \mathbb{N}}$ of $p_1$ and $p_2$ by simple functions from below, converging in $L^1(X,\mu)$ norm, such that
\begin{equation}
\int_X \sqrt{p_1\,p_2} \total \mu = \lim_{n \to \infty} \int_X \sqrt{p_1^{(n)} \, p_2^{(n)}} \total \mu .
\end{equation}
For this sequence, let $X = \bigcup_i X_i^{(n)}$ be a partition of $X$ into disjoint sets such that $p_1^{(n)}$ and $p_2^{(n)}$ are constant on each $X_i^{(n)}$, and let $\Pb_i^{(n)} \coloneq p_i^{(n)} \mu$ be the measure associated to the approximation. Then we have
\begin{equation}
\sum_i \sqrt{ \Pb_1^{(n)}(X_i^{(n)}) \, \Pb_2^{(n)}(X_i^{(n)}) } = \int_X \sqrt{p_1^{(n)} \, p_2^{(n)}} \total \mu \leq \int_X \sqrt{p_1 \, p_2} \total \mu ,
\end{equation}
and thus
\begin{equation}
\sum_i \sqrt{ \Pb_1^{(n)}(X_i^{(n)}) \, \Pb_2^{(n)}(X_i^{(n)}) } \leq \int_X \sqrt{p_1 \, p_2} \total \mu \leq \sum_i \sqrt{ \Pb_1(X_i^{(n)}) \, \Pb_2(X_i^{(n)}) } \, .
\end{equation}
The difference between the left- and right-hand side can be written as
\begin{equation}
\begin{split}
&\sum_i \abs*{ \sqrt{ \Pb_1^{(n)}(X_i^{(n)}) \, \Pb_2^{(n)}(X_i^{(n)}) } - \sqrt{ \Pb_1(X_i^{(n)}) \, \Pb_2(X_i^{(n)}) } } \\
&\quad\leq \sum_i \abs*{ \sqrt{ \Pb_1^{(n)}(X_i^{(n)}) } - \sqrt{ \Pb_1(X_i^{(n)}) } } \, \sqrt{ \Pb_2^{(n)}(X_i^{(n)}) } \\
&\qquad+ \sum_i \sqrt{ \Pb_1(X_i^{(n)}) } \, \abs*{ \sqrt{ \Pb_2^{(n)}(X_i^{(n)}) } - \sqrt{ \Pb_2(X_i^{(n)}) } } ,
\end{split}
\end{equation}
and since by construction $\Pb_i^{(n)} \leq \Pb_i$, we also have
\begin{equation}
\begin{split}
&\left[ \sqrt{ \Pb_1^{(n)}(X_i^{(n)}) } - \sqrt{ \Pb_1(X_i^{(n)}) } \right]^2 \\
&\quad= \Pb_1(X_i^{(n)}) - \Pb_1^{(n)}(X_i^{(n)}) - 2 \sqrt{ \Pb_1^{(n)}(X_i^{(n)}) } \left[ \sqrt{ \Pb_1(X_i^{(n)}) } - \sqrt{ \Pb_1^{(n)}(X_i^{(n)}) } \right] \\
&\quad\leq \Pb_1(X_i^{(n)}) - \Pb_1^{(n)}(X_i^{(n)}) .
\end{split}
\end{equation}
Using then the Cauchy--Schwarz inequality and the fact that the $\Pb_i$ are probability measures and thus bounded by 1, it follows that
\begin{equation}
\begin{split}
&\left[ \sum_i \abs*{ \sqrt{ \Pb_1^{(n)}(X_i^{(n)}) } - \sqrt{ \Pb_1(X_i^{(n)}) } } \, \sqrt{ \Pb_2^{(n)}(X_i^{(n)}) } \right]^2 \\
&\quad\leq \sum_i \left[ \sqrt{ \Pb_1^{(n)}(X_i^{(n)}) } - \sqrt{ \Pb_1(X_i^{(n)}) } \right]^2 \sum_j \Pb_2^{(n)}(X_j^{(n)}) \\
&\quad\leq \sum_i \left[ \Pb_1(X_i^{(n)}) - \Pb_1^{(n)}(X_i^{(n)}) \right] ,
\end{split}
\end{equation}
and the analogous equation with $\Pb_1$ and $\Pb_2$ exchanged. Furthermore, we obtain
\begin{equation}
\sum_i \abs*{ \Pb_1^{(n)}(X_i^{(n)}) - \Pb_1(X_i^{(n)}) } = \sum_i \int_{X_i^{(n)}} \left( p_1 - p_1^{(n)} \right) \total \mu = \norm*{ p_1 - p_1^{(n)} }_{L^1(X,\mu)} .
\end{equation}
Choosing $n$ sufficiently large, we can thus make the difference between $\int_X \sqrt{p_1 \, p_2} \total \mu$ and $\sum_i \sqrt{ \Pb_1(X_i^{(n)}) \, \Pb_2(X_i^{(n)}) }$ as small as we like.

The inequality~\eqref{eq:monotonity-ineq1} thus holds if we can show that $\sqrt{ P(\omega_1,\omega_2) } \leq \sum_i \sqrt{ \Pb_1(X_i) \, \Pb_2(X_i) }$ for any finite partition of $X$, since then it will also be smaller than the infimum. However, this follows easily from the commutant formula, Lemma~\ref{lem:commutant}. Namely, since $\sum_i E(X_i) = 1$ for any partition of $X = \bigcup_i X_i$ into disjoint subsets, for $x' \in \M'$ we have
\begin{equation}
\begin{split}
\abs{ \ip{\xi_1}{x' \xi_2} } &\leq \sum_i \abs{ \ip{\xi_1}{x' E(X_i) \xi_2} } = \sum_i \abs*{ \left\langle \sqrt{ E(X_i) } \, \xi_1, x' \sqrt{ E(X_i) } \, \xi_2 \right\rangle } \\
&\leq \sum_i \norm*{ \sqrt{ E(X_i) } \, \xi_1 } \norm{ x' } \norm*{ \sqrt{ E(X_i) } \, \xi_2 } \\
&= \norm{ x' } \sum_i \sqrt{ \omega_1(E(X_i)) \, \omega_2(E(X_i)) } = \norm{ x' } \sum_i \sqrt{ \Pb_1(X_i) \, \Pb_2(X_i) } \, ,
\end{split}
\end{equation}
where $\omega_1, \omega_2$ are realized as vector states for vectors $\xi_1, \xi_2 \in \Hs$, and we used that $x' \in \M'$ commutes with $E(X_i) \in \M$. Taking the supremum over $x'$ with $\norm{ x' } \leq 1$, by Lemma~\ref{lem:commutant} we obtain as required
\begin{equation}
\sqrt{P(\omega_1,\omega_2)} = \sup_{x' \in \M' \colon \norm{ x' } \leq 1} \abs{ \ip{\xi_1}{x'\xi_2} } \leq \sum_i \sqrt{ \Pb_1(X_i) \, \Pb_2(X_i) } .
\end{equation}
\end{proof}
For a finite-dimensional Hilbert space, the so-called Fuchs--Caves measurement~\cite{FuchsCaves1995} $E_\mathrm{FC}$ achieves equality in \eqref{eq:monotonity-ineq1}. That is, the Bures distance is the Hellinger distance of the most discriminating local measurement. For a general von Neumann algebra, no such measurement needs to exist, but we can still show that $\sqrt{P(\omega_1,\omega_2)}$ is the infimum of the Bhattacharyya coefficient over all measurements in $\M$.
\begin{lemma}[Infimum of measurements]
\label{lem:infimum}
Let $\omega_1,\omega_2$ be normal states on $\M$, and let $\Pb_j = \omega_j \circ E$ be the outcome distributions of a measurement $E \in \mathcal{E}_\M$. Then it holds that
\begin{equation}
\sqrt{P(\omega_1,\omega_2)} = \inf_{E \in \mathcal{E}_\M} B(\Pb_1,\Pb_2) ,
\end{equation}
and the infimum may be taken over projective measurements only.
\end{lemma}
\begin{proof}
By Lemma~\ref{lem:mono}, we have $\sqrt{P(\omega_1,\omega_2)} \leq \inf_E B(\Pb_1,\Pb_2)$. To show the reverse inequality, choose $0 < x \in \M$, and let $E_x$ be the spectral measure of $x$, which is a POVM with outcome space $((0,\norm{x}],\mathcal{B}((0,\norm{x}]))$ fulfilling the requirements of Def.~\ref{def:povm}, and thus a measurement. Let $\Pb_i^x = \omega_i \circ E_x$ be the corresponding outcome distributions, let $\mu_x \coloneq (\Pb_1^x + \Pb_2^x)/2$ be a dominating probability measure, and denote by $p_i^x$ the densities of $\Pb_i^x$ with respect to $\mu_x$. Using the arithmetic-geometric mean inequality, we compute
\begin{equation}
\begin{split}
B(\Pb_1^x,\Pb_2^x) &= \int_{(0,\infty)} \sqrt{ p_1^x(t) \, p_2^x(t) } \total \mu_x(t) \leq \frac{1}{2} \int_{(0,\infty)} \left[ t \, p_1^x(t) + \frac{1}{t} p_2^x(t) \right] \total \mu_x(t) \\
&= \frac{1}{2} \int_{(0,\infty)} t \total \Pb_1^x(t) + \frac{1}{2} \int_{(0,\infty)} \frac{1}{t} \total \Pb_2^x(t) = \frac{1}{2} \omega_1(x) + \frac{1}{2} \omega_2\left( x^{-1} \right) ,
\end{split}
\end{equation}
and it follows that
\begin{equation}
\inf_E B(\Pb_1,\Pb_2) \leq \inf_{x \in \M \colon x > 0} B(\Pb_1^x,\Pb_2^x) \leq \frac{1}{2} \inf_{x \in \M \colon x > 0} \left[ \omega_1(x) + \omega_2\left( x^{-1} \right) \right] = \sqrt{ P(\omega_1,\omega_2) } \, ,
\end{equation}
where the last equality is the variational formula~\cite[Cor.~2 (5)]{AlbertiUhlmann2000} that we already employed in the proof of Thm.~\ref{thm:exist}. Since $E_x$ is a projective measurement, this also shows that we may restrict the infimum to those.
\end{proof}

These results apply to any von Neumann algebra. In particular, we can apply them to the product states $\omega^{\otimes\nu}$ on the tensor product algebra $\M^{\otimes\nu}$, which correspond to a repeated simultaneous preparation of the same system $\nu$ times. A measurement $E$ in $\M^{\otimes \nu}$ corresponds to a collective measurement on the $\nu$ copies; a repeated independent measurement of each copy corresponds to $E^{\otimes \nu}$ for some measurement $E$ in the single algebra $\M$. Interestingly, it is enough to consider such measurements, since Uhlmann's fidelity is multiplicative under tensor products:
\begin{lemma}[Product states]
\label{lem:product}
For normal states $\omega_1,\omega_2$ on $\M$ and $\nu \in \mathbb{N}$, it holds that
\begin{equation}
\sqrt{ P\left( \omega_1^{\otimes\nu}, \omega_2^{\otimes\nu} \right) } = \sqrt{ P(\omega_1, \omega_2) }^{\,\nu} .
\end{equation}
\end{lemma}
\begin{proof}
We employ the second variational formula~\cite[Thm.~2 (1)]{AlbertiUhlmann2000}
\begin{equation}
\sqrt{ P(\omega_1,\omega_2) } = \inf_{x \in \M \colon x > 0} \sqrt{ \omega_1(x) \, \omega_2\left( x^{-1} \right) } ,
\end{equation}
which follows from the previous one by choosing $\lambda x$ with $\lambda = \sqrt{ \omega_2\left( x^{-1} \right) / \omega_1(x) }$ instead of $x$. For any positive $y \in \M$, the tensor product $y^{\otimes \nu}$ is a positive operator in $\M^{\otimes \nu}$, such that
\begin{equation}
\begin{split}
\sqrt{ P\left( \omega_1^{\otimes\nu}, \omega_2^{\otimes\nu} \right) } &= \inf_{x \in \M^{\otimes \nu} \colon x > 0} \sqrt{ \omega_1^{\otimes\nu}(x) \, \omega_2^{\otimes\nu}\left( x^{-1} \right) } \\
&\leq \sqrt{ \omega_1^{\otimes\nu}\left( y^{\otimes \nu} \right) \, \omega_2^{\otimes\nu}\left( (1/y)^{\otimes \nu} \right) } = \left[ \omega_1(y) \, \omega_2\left( y^{-1} \right) \right]^\frac{\nu}{2} ,
\end{split}
\end{equation}
and taking the infimum over all $0 < y \in \M$, we obtain
\begin{equation}
\sqrt{ P\left( \omega_1^{\otimes\nu}, \omega_2^{\otimes\nu} \right) } \leq \sqrt{ P(\omega_1, \omega_2) }^{\,\nu} .
\end{equation}

For the reverse inequality, we use the commutant formula, Lemma~\ref{lem:commutant}. If $\omega_1, \omega_2$ are realized as vector states for vectors $\xi_1, \xi_2 \in \Hs$, their tensor products $\omega_i^{\otimes \nu}$ are realized as vector states for vectors $\xi_i^{\otimes \nu} \in \Hs^{\otimes \nu}$. By Lemma~\ref{lem:commutant}, we have
\begin{equation}
\begin{split}
\sqrt{ P(\omega_1,\omega_2) }^{\, \nu} &= \sup_{y' \in \M' \colon \norm{y'} \leq 1} \abs{ \ip{\xi_1}{y' \xi_2} }^\nu = \sup_{y' \in \M' \colon \norm{y'} \leq 1} \abs{ \ip{\xi_1^{\otimes \nu}}{(y')^{\otimes \nu} \xi_2^{\otimes \nu}} } \\
&\leq \sup_{x' \in (\M')^{\otimes \nu} \colon \norm{x'} \leq 1} \abs{ \ip{\xi_1^{\otimes \nu}}{x' \xi_2^{\otimes \nu}} } = \sqrt{ P\left( \omega_1^{\otimes\nu}, \omega_2^{\otimes\nu} \right) } ,
\end{split}
\end{equation}
since elements $(y')^{\otimes \nu} \in (\M')^{\otimes \nu}$ with $\norm{ y' } \leq 1$ are a subset of all elements $x' \in (\M')^{\otimes \nu}$ with norm bounded by 1, and the last equality is Lemma~\ref{lem:commutant} for $\M^{\otimes \nu}$ together with the commutation theorem $(\M^{\otimes \nu})' = (\M')^{\otimes \nu}$~\cite[Thm.~11.2.16]{KadisonRingrose1986}.
\end{proof}

\subsection{Fisher information and the quantum Cram\'er--Rao bound}

In analogy with Def.~\ref{def:sld}, we define the \emph{classical Fisher information} of the measurement $E$ through the outcome distribution $\Pb_\theta = \omega_\theta \circ E$ as
\begin{equation}
\label{eq:fcl-def}
F^\mathrm{cl}(\theta;E) \coloneq 8 \lim_{\varepsilon \to 0} \frac{1 - B(\Pb_\theta,\Pb_{\theta+\varepsilon})}{\varepsilon^2}
\end{equation}
whenever the limit exists. If $\theta \mapsto \sqrt{ p_\theta }$ is differentiable in $L^2(X,\mu)$ (differentiability in quadratic mean~\cite[Ch.~7]{VanDerVaart1998}), the last equality of \eqref{eq:bhatcoef} yields
\begin{equation}
\label{eq:fcl-expression}
F^\mathrm{cl}(\theta;E) = 4 \lim_{\varepsilon \to 0} \frac{\norm{ \sqrt{p_{\theta+\varepsilon}} - \sqrt{p_\theta} }^2_{L^2(X,\mu)}}{\varepsilon^2} = 4 \, \norm{ \partial_\theta \sqrt{p_\theta} }^2_{L^2(X,\mu)} ,
\end{equation}
which for smooth densities is the usual Fisher information $\int_X (\partial_\theta p_\theta)^2 p_\theta^{-1} \total \mu$. We then obtain
\begin{lemma}[Fisher information bound]
\label{lem:fisher-bound}
Let $\theta \mapsto \omega_\theta$ be a curve of normal states on a von Neumann algebra $\M$ for which $\Fs_\M(\theta)$ exists, and let $E \in \mathcal{E}_{\M^{\otimes \nu}}$ be a measurement. If the classical Fisher information $F^\mathrm{cl}(\theta; E)$ of the outcome distributions $\omega_\theta^{\otimes\nu} \circ E$ exists, the following inequality holds:
\begin{equation}
F^\mathrm{cl}(\theta; E) \leq \nu \, \Fs_\M(\theta) .
\end{equation}
\end{lemma}
\begin{proof}
By Lemmas~\ref{lem:product} and \ref{lem:infimum}, we have
\begin{equation}
\sqrt{ P(\omega_1, \omega_2) }^{\,\nu} = \sqrt{ P\left( \omega_1^{\otimes\nu}, \omega_2^{\otimes\nu} \right) } = \inf_E B(\Pb_1^\nu,\Pb_2^\nu)
\end{equation}
with the outcome distributions $\Pb_i^\nu = \omega_i^{\otimes \nu} \circ E$ and measurements $E$ in $\M^{\otimes\nu}$. Choosing $\omega_1 = \omega_\theta$ and $\omega_2 = \omega_{\theta+\varepsilon}$, we obtain using Def.~\ref{def:sld}
\begin{equation}
\label{eq:sld-tensor}
\Fs_{\M^{\otimes \nu}}(\theta) = 8 \lim_{\varepsilon \to 0} \frac{1 - \sqrt{ P\left( \omega_\theta^{\otimes\nu}, \omega_{\theta+\varepsilon}^{\otimes\nu} \right) }}{\varepsilon^2} = 8 \lim_{\varepsilon \to 0} \frac{1 - \sqrt{ P(\omega_\theta, \omega_{\theta+\varepsilon}) }^{\,\nu}}{\varepsilon^2} = \nu \, \Fs_{\M}(\theta) .
\end{equation}
Hence for any $E$ we have
\begin{equation}
\nu \, \Fs_{\M}(\theta) \geq 8 \lim_{\varepsilon \to 0} \frac{1 - B(\Pb_1^\nu,\Pb_2^\nu)}{\varepsilon^2} = F^\mathrm{cl}(\theta; E) .
\end{equation}
\end{proof}
Under stronger assumptions, the quantum Fisher information can be obtained as a supremum over measurements of the classical Fisher information.
\begin{lemma}[Quantum Fisher information]
\label{lem:fisher-bound-supremum}
Let $\theta \mapsto \omega_\theta$ be a curve of faithful normal states on a von Neumann algebra $\M$ for which $\theta \mapsto \xi_\theta$ is continuously differentiable in norm, where $\xi_\theta$ is the canonical implementing vector of $\omega_\theta$. Then it holds that
\begin{equation}
\Fs_\M(\theta) = \frac{1}{\nu} \sup_{E \in \mathcal{E}_{\M^{\otimes \nu}}} F^\mathrm{cl}(\theta; E) ,
\end{equation}
where the supremum may be restricted to projective measurements.
\end{lemma}
\begin{proof}
By Thm.~\ref{thm:exist} and Cor.~\ref{cor:equivforms}, $\Fs_\M(\theta)$ exists for the given curve and is given by
\begin{equation}
\Fs_\M(\theta) = \sup_{x = x^* \in \M \colon \Var_{\omega_\theta}(x)>0} \frac{\dot\omega_\theta(x)^2}{\Var_{\omega_\theta}(x)} .
\end{equation}
Since neither $\dot\omega_\theta(x)$ nor $\Var_{\omega_\theta}(x)$ change if a multiple of the identity is added to $x$, it suffices to consider $x > 0$. By functional calculus, for each $0 < x = x^* \in \M$ we find a sequence $(x_n)_{n \in \mathbb{N}}$ with $x_n = \sum_{k=1}^{K(n)} \lambda_k^{(n)} e_k^{(n)}$, where $\lambda_k^{(n)} \geq 0$ with $\lambda_k^{(n)} > 0$ for at least one $k$, and the $e_k^{(n)} \in \M$ are pairwise orthogonal spectral projections with $\sum_{k=1}^{K(n)} e_k^{(n)} = 1$~\cite[Sec.~5.2]{KadisonRingrose1997}. These projections define a projective measurement $E_n$ in $\M$ with outcome space $\bigl( \{1,\ldots,K(n)\}, 2^{\{1,\ldots,K(n)\}} \bigr)$ via $E_n(\{k\}) \coloneq e_k^{(n)}$, and the corresponding classical Fisher information $F^\mathrm{cl}(\theta;E_n)$ can be obtained from \eqref{eq:fcl-expression}:
\begin{equation}
F^\mathrm{cl}(\theta;E_n) = \norm*{ \frac{1}{\sqrt{p_\theta}} \partial_\theta p_\theta }^2_{L^2(X,\mu)} = \sum_{k=1}^{K(n)} \frac{\Bigl[ \partial_\theta \omega_\theta\left( E_n(\{k\}) \right) \Bigr]^2}{\omega_\theta\left( E_n(\{k\}) \right)} = \sum_{k=1}^{K(n)} \frac{\dot\omega_\theta\left( e_k^{(n)} \right)^2}{\omega_\theta\left( e_k^{(n)} \right)} .
\end{equation}
On the other hand, we have
\begin{equation}
\begin{split}
\dot\omega_\theta(x_n)^2 &= \left[ \sum_{k=1}^{K(n)} \lambda_k^{(n)} \dot\omega_\theta\left( e_k^{(n)} \right) \right]^2 = \left[ \sum_{k=1}^{K(n)} \left( \lambda_k^{(n)} - \omega_\theta(x_n) \right) \dot\omega_\theta\left( e_k^{(n)} \right) \right]^2 \\
&\leq \left[ \sum_{k=1}^{K(n)} \left( \lambda_k^{(n)} - \omega_\theta(x_n) \right)^2 \omega_\theta\left( e_k^{(n)} \right) \right] \sum_{j=1}^{K(n)} \frac{\dot\omega_\theta\left( e_j^{(n)} \right)^2}{\omega_\theta\left( e_j^{(n)} \right)} ,
\end{split}
\end{equation}
using that $\sum_{k=1}^{K(n)} \dot\omega_\theta\left( e_k^{(n)} \right) = \dot\omega_\theta(1) = 0$ and the Cauchy--Schwarz inequality. Since
\begin{equation}
\Var_{\omega_\theta}(x_n) = \omega_\theta\left( \left( x_n - \omega_\theta(x_n) \right)^2 \right) = \sum_{k=1}^{K(n)} \left( \lambda_k^{(n)} - \omega_\theta(x_n) \right)^2 \omega_\theta\left( e_k^{(n)} \right) ,
\end{equation}
we thus obtain
\begin{equation}
\frac{\dot\omega_\theta(x_n)^2}{\Var_{\omega_\theta}(x_n)} \leq \sum_{j=1}^{K(n)} \frac{\dot\omega_\theta\left( e_j^{(n)} \right)^2}{\omega_\theta\left( e_j^{(n)} \right)} = F^\mathrm{cl}(\theta;E_n) \leq \sup_{E \in \mathcal{E}_\M} F^\mathrm{cl}(\theta;E) ,
\end{equation}
where the supremum may be restricted to projective measurements. Taking first the limit $n \to \infty$ and then the supremum over $x$ yields
\begin{equation}
\Fs_\M(\theta) \leq \sup_{E \in \mathcal{E}_\M} F^\mathrm{cl}(\theta;E) .
\end{equation}
Using the additivity of the quantum Fisher information~\eqref{eq:sld-tensor}, this inequality for the curve $\omega_\theta^{\otimes\nu}$ on $\M^{\otimes \nu}$ (which satisfies the same assumptions, with implementing vectors $\xi_\theta^{\otimes\nu}$) and finally Lemma~\ref{lem:fisher-bound}, it follows that
\begin{equation}
\nu \, \Fs_\M(\theta) = \Fs_{\M^{\otimes \nu}}(\theta) \leq \sup_{E \in \mathcal{E}_{\M^{\otimes \nu}}} F^\mathrm{cl}(\theta;E) \leq \nu \, \Fs_\M(\theta) ,
\end{equation}
and thus equality.
\end{proof}

We first recall the classical Cram\'er--Rao bound for locally unbiased estimators, whose proof is so simple that we reproduce it here.
\begin{theorem}[Classical Cram\'er--Rao bound]
\label{thm:classical-cramer-rao}
Let $(X,\Sigma)$ be a measure space, let $\Pb_\theta$ with $\theta \in \mathbb{R}$ be a family of probability measures on $(X,\Sigma)$ which for $\theta$ in an open interval around a fixed $\theta_0$ have densities $p_\theta$ with respect to a $\sigma$-finite measure $\mu$, and assume that $p_\theta$ is differentiable at $\theta_0$ for $\mu$-almost every $x \in X$. Then every estimator $\hat\theta \colon X \to \mathbb{R}$ that is $\Sigma$-measurable, regular and locally unbiased at $\theta_0$,
\begin{align}
\int_X \hat{\theta}(x) p_{\theta_0}(x) \total \mu(x) &= \theta_0 , \\
\int_S \left[ \hat{\theta}(x) - \theta_0 \right] \dot p_{\theta_0}(x) \total \mu(x) &= 1 , \ S \coloneq \{x \in X \colon p_{\theta_0}(x) > 0\} \label{eq:classical-cramer-rao-regularity}
\end{align}
satisfies
\begin{equation}
\Var_{\theta_0}(\hat\theta) \geq \frac{1}{F^\mathrm{cl}(\theta_0)}
\end{equation}
with the Fisher information
\begin{equation}
\label{eq:classical-fisher-cramer-rao}
F^{\mathrm{cl}}(\theta_0) \coloneq \int_S \frac{\dot p_{\theta_0}(x)^2}{p_{\theta_0}(x)} \total \mu(x) \geq 0 .
\end{equation}
\end{theorem}
\begin{proof}
If $\Var_{\theta_0}(\hat\theta) = \infty$ or $F^{\mathrm{cl}}(\theta_0) = \infty$, the inequality holds trivially. If $F^{\mathrm{cl}}(\theta_0) = 0$, we have $\dot p_{\theta_0}(x) = 0$ for $\mu$-almost every $x \in S$, which contradicts the regularity assumption~\eqref{eq:classical-cramer-rao-regularity}. In all other cases, the Cauchy--Schwarz inequality shows that
\begin{equation}
\begin{split}
1 &= \abs*{ \int_S \left[ \hat{\theta}(x) - \theta_0 \right] \sqrt{ p_{\theta_0}(x) } \, \frac{\dot p_{\theta_0}(x)}{\sqrt{ p_{\theta_0}(x) }} \total \mu(x) }^2 \\
&\leq \int_S \left[ \hat{\theta}(x) - \theta_0 \right]^2 p_{\theta_0}(x) \total \mu(x) \, \int_S \frac{\dot p_{\theta_0}(x)^2}{p_{\theta_0}(x)} \total \mu(x) = \Var_{\theta_0}(\hat\theta) F^{\mathrm{cl}}(\theta_0) .
\end{split}
\end{equation}
As explained around Eq.~\eqref{eq:fcl-expression}, equality of the classical Fisher information defined through \eqref{eq:fcl-def} with the expression~\eqref{eq:classical-fisher-cramer-rao} holds if $\theta \mapsto \sqrt{ p_\theta }$ is moreover differentiable in quadratic mean.
\end{proof}

We then obtain our central result:
\begin{theorem}[Quantum Cram\'er--Rao bound]
\label{thm:qcrb}
Let $\theta \mapsto \omega_\theta$ be a curve of normal states on a von Neumann algebra $\M$ for which $\Fs_\M(\theta)$ exists, and let $E$ be a measurement in $\M^{\otimes\nu}$. Assume that there is $\theta_0$ such that the densities $p_\theta$ of the outcome distributions $\Pb_\theta \coloneq \omega_\theta^{\otimes\nu} \circ E$ are differentiable in quadratic mean at $\theta_0$, such that the classical Fisher information $F^\mathrm{cl}(\theta_0; E)$ exists and is given by~\eqref{eq:fcl-expression}. Under the regularity assumptions of the classical Cram\'er--Rao bound (Thm.~\ref{thm:classical-cramer-rao}), every estimator $\hat\theta \colon X \to \mathbb{R}$ that is locally unbiased at $\theta_0$ satisfies
\begin{equation}
\Var_{\theta_0}(\hat\theta) \geq \frac{1}{F^\mathrm{cl}(\theta_0; E)} \geq \frac{1}{\nu\,\Fs_\M(\theta_0)} .
\end{equation}
\end{theorem}
\begin{proof}
By differentiability in quadratic mean, the $L^2$ derivative of $\sqrt{p_\theta}$ at $\theta_0$ is equal to $\dot p_{\theta_0}/(2 \sqrt{p_{\theta_0}})$ almost everywhere on $S$, and it vanishes almost everywhere on $X \setminus S$, since the difference quotients $\sqrt{p_{\theta_0+\varepsilon}}/\varepsilon$ are non-negative there for $\varepsilon > 0$ and non-positive for $\varepsilon < 0$. Hence $F^\mathrm{cl}(\theta_0; E)$ coincides with the Fisher information $F^\mathrm{cl}(\theta_0)$ of Thm.~\ref{thm:classical-cramer-rao}. The first inequality is then the classical Cram\'er--Rao bound (Thm.~\ref{thm:classical-cramer-rao}), and the second follows from Lemma~\ref{lem:fisher-bound}.
\end{proof}

Under stronger assumptions the bound is sharp: its right-hand side is the infimum of the variance over all (possibly collective) measurements on $\nu$ copies and all admissible estimators. In fact, one may restrict to local and projective measurements: collective measurements give no advantage.
\begin{theorem}[Sharpness of the quantum Cram\'er--Rao bound]
\label{thm:qcrb_inf}
Let $\theta \mapsto \omega_\theta$ be a curve of faithful normal states on a von Neumann algebra $\M$ for which $\theta \mapsto \xi_\theta$ is continuously differentiable in norm, where $\xi_\theta$ is the canonical implementing vector of $\omega_\theta$, and assume that $\Fs_\M(\theta_0) > 0$. Then it holds that
\begin{equation}
\inf_{E,\hat\theta} \Var_{\theta_0}(\hat\theta) = \frac{1}{\nu \, \Fs_\M(\theta_0)} ,
\end{equation}
where the infimum runs over all measurements $E \in \mathcal{E}_{\M^{\otimes \nu}}$ and all estimators $\hat\theta$ satisfying the assumptions of Thm.~\ref{thm:qcrb}. The infimum may be restricted to measurements $E = E_1^{\otimes \nu}$, where $E_1 \in \mathcal{E}_\M$ is a projective measurement with finitely many outcomes.
\end{theorem}
\begin{proof}
By Thm.~\ref{thm:qcrb}, we have $\Var_{\theta_0}(\hat\theta) \geq 1/(\nu \, \Fs_\M(\theta_0))$ for every measurement $E$ and every estimator $\hat{\theta}$. For the converse, let $E_1 \in \mathcal{E}_\M$ be a projective measurement with finitely many outcomes, all of which have nonzero projections, satisfying $F^\mathrm{cl}(\theta_0;E_1) > 0$. Since $\omega_{\theta_0}$ is faithful, all outcomes of $E_1$ and hence of $E \coloneq E_1^{\otimes\nu}$ have positive probability $p_{\theta_0}(k) > 0$, and since the outcome distribution of $E$ is a product, $F^\mathrm{cl}(\theta_0;E) = \nu F^\mathrm{cl}(\theta_0;E_1)$. We then define the estimator
\begin{equation}
\hat\theta(k) \coloneq \theta_0 + \frac{\dot p_{\theta_0}(k)}{p_{\theta_0}(k) \, F^\mathrm{cl}(\theta_0;E)} ,
\end{equation}
which is regular and locally unbiased at $\theta_0$, and has variance $\Var_{\theta_0}(\hat\theta) = 1/F^\mathrm{cl}(\theta_0;E) = 1/(\nu F^\mathrm{cl}(\theta_0;E_1))$. By Lemma~\ref{lem:fisher-bound-supremum} with $\nu = 1$, the supremum of $F^\mathrm{cl}(\theta_0;E_1)$ over such $E_1$ is $\Fs_\M(\theta_0)$.
\end{proof}

\section{Optimal measurements and achievement}
\label{sec:attain}

Thm.~\ref{thm:qcrb} bounds the classical Fisher information of every measurement on $\nu$ copies by $\nu\,\Fs_\M(\theta)$, and Thm.~\ref{thm:qcrb_inf} shows that the bound is sharp for every curve $\theta \mapsto \omega_\theta$ of faithful states whose canonical implementing vectors are continuously differentiable in norm. That is, even if there does not exist an optimal measurement that exactly achieves the bound, for every $\delta > 0$ one may find a measurement $E_\delta \in \mathcal{E}_{\M^{\otimes \nu}}$ such that the difference between $F^\mathrm{cl}(\theta_0; E_\delta)$ and $\nu\,\Fs_\M(\theta_0)$ is less than $\delta$, by the definition of the supremum. In finite dimensions, an optimal measurement exists and consists of the eigenprojections of the SLD, which are measurements in the algebra. However, a general von Neumann algebra $\M$ may be a type III factor, which has no trace and no minimal projections (equivalently, no pure normal states), and thus it is not clear how an SLD can be defined. Moreover, even if the SLD and its spectral projections can be defined, they do not need to lie in $\M$. Nevertheless, we will show that whenever a SLD can be defined as an operator affiliated with $\M$, its spectral measure gives an optimal measurement achieving the bound (Thm.~\ref{thm:optimal}).

Why is it important that the bound can be achieved, at least arbitrarily close? An upper bound alone does not single out $\Fs_\M$, since every larger quantity is also an upper bound. Braunstein and Caves make the same point in finite dimensions: ``an unachievable upper bound cannot be used to define statistical distance''~\cite[p.~3442]{BraunsteinCaves1994}.

The first reason to consider $\Fs_\M$ is its operational meaning: If the bound is sharp, $\Fs_\M(\theta)$ is the best precision available to an observer who can measure only in $\M$, i.e., in the region in which the observables in $\M$ are localized. This turns comparisons of Fisher informations into comparisons of achievable precisions. For a von Neumann subalgebra $\mathcal{N} \subset \M$, for instance the algebra associated to a smaller region, the inclusion $\mathcal{N} \hookrightarrow \M$ is unital and completely positive, and \cite[Cor.~4]{Alberti1983} shows that Uhlmann's fidelity cannot decrease under restriction: $P(\omega_\theta\rvert_{\mathcal N},\omega_{\theta+\varepsilon}\rvert_{\mathcal N}) \geq P(\omega_\theta,\omega_{\theta+\varepsilon})$. Hence, whenever the limits defining the quantum Fisher information exist, we have $\Fs_{\mathcal N}(\theta) \leq \Fs_\M(\theta)$: a smaller region can only lose information about the family of states $\omega_\theta$.

Similarly, our results quantify the loss of a local observer relative to a global one: the quantum Fisher information that a local observer can access is only the projection on the standard subspace associated to the local region, see Example~\ref{ex:qubits} and Sec.~\ref{sec:wedge} for explicit expressions of the loss. If the bound were not sharp, all these statements would only compare two bounds, and a gap between them could be an artifact of a loose bound.

In finite dimensions, every monotone metric on the state space with the classical normalization bounds the classical Fisher information of every measurement, and the SLD metric is the smallest of them~\cite{Petz1996}. Each such metric therefore yields a valid Cram\'er--Rao bound. An example is the Bogoliubov--Kubo--Mori (BKM) metric, the Hessian of the relative entropy, which on type III von Neumann algebras is defined through the Araki--Uhlmann relative entropy~\cite{Araki1975,Uhlmann1977} and has recently been evaluated for local states in quantum field theory~\cite{Chatterjee2026}. We show that the BKM Fisher information is never smaller than $\Fs_\M$ (Prop.~\ref{prop:bkm}). Since no measurement in $\M$ has classical Fisher information larger than $\Fs_\M$ (Lemma~\ref{lem:fisher-bound}), a bound built on a metric that exceeds $\Fs_\M$ at some $\theta$ cannot be approached there by any measurement in the region. Among these metrics, $\Fs_\M$ gives a bound that is achieved (Lemma~\ref{lem:fisher-bound-supremum} and Thms.~\ref{thm:qcrb}, \ref{thm:qcrb_inf}).

\subsection{Unbounded observables and the SLD}

Since in finite dimensions the optimal measurement is given by eigenprojections of the SLD, we try to extend the SLD and its projections to general von Neumann algebras.

\begin{definition}[Symmetric logarithmic derivative]
\label{def:sldop}
Let $\theta \mapsto \omega_\theta$ be a curve of faithful normal states on a von Neumann algebra $\M$ for which $\theta \mapsto \xi_\theta$ is continuously differentiable in norm, where $\xi_\theta$ is the canonical implementing vector of $\omega_\theta$. A self-adjoint operator $L_\theta$ affiliated with $\M$, and with $\xi_\theta \in D(L_\theta)$, is a \emph{symmetric logarithmic derivative} (SLD) of the curve at $\theta$ if
\begin{equation}
L_\theta \xi_\theta = 2 \Pi_\theta \dot\xi_\theta ,
\end{equation}
with $\Pi_\theta$ the real-orthogonal projection defined in Thm.~\ref{thm:exist}.
\end{definition}
For the quantum-mechanical case with $\M = B(\Hs) \otimes 1$ acting on the bipartite Hilbert space $\Hs \otimes \Hs$, this is the usual SLD as shown after Cor.~\ref{cor:equivforms}, see Eq.~\eqref{eq:SLD-qm-projection}. In this case, its eigenprojections give an optimal measurement, and we want to prove the same in the general case.  $L_\theta$ will generally be unbounded, and   therefore we first need to extend our results to unbounded operators affiliated with $\M$.

Let thus $L$ be self-adjoint and affiliated with $\M$. Its spectral projections lie in $\M$, so its spectral measure $E^L$ is a measurement in $\M$ with outcome space $(\mathbb{R},\mathcal{B}(\mathbb{R}))$ by Def.~\ref{def:povm}. If $\xi_\theta \in D(L)$, spectral calculus shows that the mean and variance of the outcome distribution of $E^L$ are given by $\omega_\theta(L) = \ip{\xi_\theta}{L \xi_\theta}$ and $\Var_{\omega_\theta}(L) = \norm{ \left( L - \omega_\theta(L) \right) \xi_\theta }^2$.
\begin{theorem}[Optimal SLD measurement]
\label{thm:optimal}
Let $L_\theta$ be a SLD with $\xi_{\theta+\varepsilon} \in D(L_\theta)$ for all sufficiently small $\abs{\varepsilon}$, and assume that $\varepsilon \mapsto L_\theta \xi_{\theta+\varepsilon}$ is norm-continuous at $\varepsilon = 0$. Then the classical Fisher information $F^\mathrm{cl}(\theta;E^{L_\theta})$ exists and equals $\Fs_\M(\theta)$.
\end{theorem}
\begin{proof}
Since $L_\theta$ is self-adjoint, we obtain
\begin{equation}
\frac{\omega_{\theta+\varepsilon}(L_\theta) - \omega_\theta(L_\theta)}{\varepsilon} = \left\langle \frac{\xi_{\theta+\varepsilon} - \xi_\theta}{\varepsilon}, L_\theta \xi_{\theta+\varepsilon} \right\rangle + \left\langle L_\theta \xi_\theta, \frac{\xi_{\theta+\varepsilon} - \xi_\theta}{\varepsilon} \right\rangle ,
\end{equation}
and the limit $\varepsilon \to 0$ exists since $\theta \mapsto \xi_\theta$ is differentiable in norm. It follows that
\begin{equation}
\dot \omega_\theta(L_\theta) = \ip{\dot\xi_\theta}{L_\theta \xi_\theta} + \ip{L_\theta \xi_\theta}{\dot\xi_\theta} = 2 \Re \ip{\dot\xi_\theta}{L_\theta \xi_\theta} = 4 \Re \ip{\dot\xi_\theta}{ \Pi_\theta \dot\xi_\theta } = 4 \norm{ \Pi_\theta \dot\xi_\theta }^2 ,
\end{equation}
where we used the definition~\ref{def:sldop} of the SLD and that $\Pi_\theta$ is a real-orthogonal projection. By Thm.~\ref{thm:exist}, we thus obtain $\dot \omega_\theta(L_\theta) = \Fs_\M(\theta)$. Similarly, we obtain
\begin{equation}
\omega_\theta(L_\theta) = \ip{\xi_\theta}{L_\theta \xi_\theta} = \Re \ip{\xi_\theta}{L_\theta \xi_\theta} = \Re \ip{\xi_\theta}{2 \Pi_\theta \dot\xi_\theta} = 2 \Re \ip{\Pi_\theta \xi_\theta}{\dot\xi_\theta} = 0 ,
\end{equation}
where the last equality follows because $\Pi_\theta \xi_\theta = \xi_\theta$ and $2 \Re \ip{\xi_\theta}{\dot\xi_\theta} = \partial_\theta \norm{ \xi_\theta }^2 = 0$. Therefore, we also have
\begin{equation}
\label{eq:var-ltheta}
\Var_{\omega_\theta}(L_\theta) = \norm{ L_\theta \xi_\theta }^2 - \omega_\theta(L_\theta)^2 = 4 \norm{ \Pi_\theta \dot\xi_\theta }^2 = \Fs_\M(\theta) .
\end{equation}

The classical Fisher information $F^\mathrm{cl}(\theta;E^{L_\theta})$ is defined by~\eqref{eq:fcl-def}
\begin{equation}
F^\mathrm{cl}(\theta;E) = 8 \lim_{\varepsilon \to 0} \frac{1 - B(\Pb_\theta,\Pb_{\theta+\varepsilon})}{\varepsilon^2}
\end{equation}
whenever the limit exists. Let thus $\Pb_{\theta+\varepsilon} \coloneq \omega_{\theta+\varepsilon} \circ E^{L_\theta}$ be the outcome distribution of $L_\theta$ in $\omega_{\theta+\varepsilon}$, which is a probability measure on $(\mathbb{R},\mathcal{B}(\mathbb{R}))$. Let $\mu \coloneq (\Pb_\theta + \Pb_{\theta+\varepsilon})/2$ be a dominating measure, and let $p_\theta$ be the density of $\Pb_\theta$ with respect to $\mu$. Writing
\begin{equation}
\begin{split}
\omega_{\theta+\varepsilon}(L_\theta) - \omega_\theta(L_\theta) &= \int_\mathbb{R} x \total \Pb_{\theta+\varepsilon}(x) - \int_\mathbb{R} x \total \Pb_\theta(x) = \int_\mathbb{R} x \bigl[ p_{\theta+\varepsilon}(x) - p_\theta(x) \bigr] \total \mu(x) \\
&= \int_\mathbb{R} x \left[ \sqrt{ p_{\theta+\varepsilon}(x) } - \sqrt{ p_\theta(x) } \right] \left[ \sqrt{ p_{\theta+\varepsilon}(x) } + \sqrt{ p_\theta(x) } \right] \total \mu(x) ,
\end{split}
\end{equation}
we apply the Cauchy--Schwarz inequality to obtain
\begin{equation}
\bigl[ \omega_{\theta+\varepsilon}(L_\theta) - \omega_\theta(L_\theta) \bigr]^2 \leq \int_\mathbb{R} \left[ \sqrt{ p_{\theta+\varepsilon}(x) } - \sqrt{ p_\theta(x) } \right]^2 \total \mu(x) \int_\mathbb{R} x^2 \left[ \sqrt{ p_{\theta+\varepsilon}(x) } + \sqrt{ p_\theta(x) } \right]^2 \total \mu(x) .
\end{equation}
The first integral is equal to $2 - 2 B(\Pb_\theta,\Pb_{\theta+\varepsilon})$ by definition of the Bhattacharyya coefficient~\eqref{eq:bhatcoef}, and for the second we use the inequality $(a+b)^2 \leq 2 a^2 + 2 b^2$ to obtain
\begin{equation}
\begin{split}
\bigl[ \omega_{\theta+\varepsilon}(L_\theta) - \omega_\theta(L_\theta) \bigr]^2 &\leq 4 \left[ 1 - B(\Pb_\theta,\Pb_{\theta+\varepsilon}) \right] \int_\mathbb{R} x^2 \bigl[ p_{\theta+\varepsilon}(x) + p_\theta(x) \bigr] \total \mu(x) \\
&= 4 \left[ 1 - B(\Pb_\theta,\Pb_{\theta+\varepsilon}) \right] \left[ \omega_{\theta+\varepsilon}\left( L_\theta^2 \right) + \omega_\theta\left( L_\theta^2 \right) \right] .
\end{split}
\end{equation}
Dividing by $\varepsilon^2$ and taking the limit $\varepsilon \to 0$, the limit of the left-hand side exists by the previous computations and equals $\dot \omega_\theta(L_\theta)^2 = \Fs_\M(\theta)^2$. Also the limit $\varepsilon \to 0$ of the last terms on the right-hand side exists and equals $2 \Var_{\omega_\theta}(L_\theta) = 2 \Fs_\M(\theta)$ by Eq.~\eqref{eq:var-ltheta}. We thus obtain
\begin{equation}
4 \liminf_{\varepsilon \to 0} \frac{1 - B(\Pb_\theta,\Pb_{\theta+\varepsilon})}{\varepsilon^2} \geq \frac{\dot \omega_\theta(L_\theta)^2}{2 \Var_{\omega_\theta}(L_\theta)} = \frac{1}{2} \Fs_\M(\theta) ,
\end{equation}
and since the proof of Lemma~\ref{lem:fisher-bound} (with $\nu = 1$) shows that
\begin{equation}
8 \limsup_{\varepsilon \to 0} \frac{1 - B(\Pb_\theta,\Pb_{\theta+\varepsilon})}{\varepsilon^2} \leq \Fs_\M(\theta)
\end{equation}
whenever $\Fs_\M(\theta)$ exists, we obtain
\begin{equation}
\Fs_\M(\theta) = 8 \lim_{\varepsilon \to 0} \frac{1 - B(\Pb_\theta,\Pb_{\theta+\varepsilon})}{\varepsilon^2} = F^\mathrm{cl}(\theta;E^{L_\theta}) .
\end{equation}
\end{proof}
Therefore, whenever a SLD can be defined which contains the implementing vectors in its domain, its spectral measure is an optimal measurement that achieves the quantum Cram\'er--Rao bound.

However, a sharp measurement of an observable is an idealization, since a real read-out adds noise. To model this, we also consider \emph{Gaussian-modulated measurements}~\cite[Sec.~3.1]{MandryschNavascues2025}. For self-adjoint $L$ affiliated with $\M$ and $\tau > 0$, the measurement $E^{L,\tau}$ is defined by
\begin{equation}
\label{eq:gauss-measurement}
\begin{split}
E^{L,\tau}(X) &\coloneq \frac{1}{\sqrt{2 \pi \tau^2}} \int_X \exp\left[ -\frac{(y-L)^2}{2\tau^2}\right] \total y \\
&= \frac{1}{\sqrt{2 \pi \tau^2}} \int_X \int_\mathbb{R} \exp\left[ - \frac{(y-t)^2}{2\tau^2} \right] \total E^L(t) \total y
\end{split}
\end{equation}
for $X \in \mathcal{B}(\mathbb{R})$, with the exponential defined by spectral calculus. All effects $E^{L,\tau}(X)$ lie in $\M$, so $E^{L,\tau}$ is a measurement in $\M$. Its outcome distribution $\Pb_{L,\tau} \coloneq \omega \circ E^{L,\tau}$ in a normal state $\omega$ is the convolution of the outcome distribution $\omega \circ E^L$ of the sharp measurement with the centered normal distribution of variance $\tau^2$. Hence, for every $c \in \mathbb{R}$ and whenever the right-hand sides are finite,
\begin{equation}
\label{eq:gauss-measurement-mean-variance}
\int_\mathbb{R} y \total \Pb_{L,\tau}(y) = \omega(L), \qquad \int_\mathbb{R} (y-c)^2 \total \Pb_{L,\tau}(y) = \omega\left( (L-c)^2 \right) + \tau^2 .
\end{equation}

We now show that Gaussian-modulated measurements of observables in $\M$ approach the quantum Cram\'er--Rao bound as the noise width tends to zero.
\begin{corollary}[Gaussian-modulated measurements]
\label{cor:gauss}
Let $\theta \mapsto \omega_\theta$ be a curve of faithful normal states on a von Neumann algebra $\M$ satisfying the assumptions of Thm.~\ref{thm:exist}. For $x = x^* \in \M$ and $\tau > 0$, let $\Pb_{\theta} \coloneq \omega_{\theta} \circ E^{x,\tau}$. Then it holds that
\begin{equation}
\label{eq:gauss-bound}
\Fs_\M(\theta) \geq 8 \limsup_{\varepsilon \to 0} \frac{1 - B(\Pb_\theta,\Pb_{\theta+\varepsilon})}{\varepsilon^2} \geq 8 \liminf_{\varepsilon \to 0} \frac{1 - B(\Pb_\theta,\Pb_{\theta+\varepsilon})}{\varepsilon^2} \geq \frac{\dot\omega_\theta(x)^2}{\Var_{\omega_\theta}(x) + \tau^2} ,
\end{equation}
and in particular, whenever $F^\mathrm{cl}(\theta;E^{x,\tau})$ exists it lies between the two outer terms. Moreover, the supremum of the right-hand side over all $x = x^* \in \M$ and all $\tau > 0$ equals $\Fs_\M(\theta)$.

The last inequality of~\eqref{eq:gauss-bound} also holds for a self-adjoint $L$ affiliated with $\M$ in place of $x$, provided that $\xi_{\theta+\varepsilon} \in D(L)$ for all sufficiently small $\abs{\varepsilon}$ and $\varepsilon \mapsto L \xi_{\theta+\varepsilon}$ is norm-continuous at $\varepsilon = 0$.
\end{corollary}
\begin{proof}
The first inequality is shown in the proof of Lemma~\ref{lem:fisher-bound} (with $\nu = 1$), which only uses that $E^{x,\tau}$ is a measurement in $\M$. For the last one, set $c \coloneq \omega_\theta(x)$, let $\mu \coloneq (\Pb_\theta + \Pb_{\theta+\varepsilon})/2$ and let $p_\theta, p_{\theta+\varepsilon}$ be the densities with respect to $\mu$. Since $\int_\mathbb{R} ( p_{\theta+\varepsilon} - p_\theta ) \total \mu = 0$, the first relation of~\eqref{eq:gauss-measurement-mean-variance} gives
\begin{equation}
\omega_{\theta+\varepsilon}(x) - \omega_\theta(x) = \int_\mathbb{R} (y-c) \left[ \sqrt{ p_{\theta+\varepsilon}(y) } - \sqrt{ p_\theta(y) } \right] \left[ \sqrt{ p_{\theta+\varepsilon}(y) } + \sqrt{ p_\theta(y) } \right] \total \mu(y) ,
\end{equation}
and the Cauchy--Schwarz inequality together with $(a+b)^2 \leq 2 a^2 + 2 b^2$ and the second relation of~\eqref{eq:gauss-measurement-mean-variance} yields, exactly as in the proof of Thm.~\ref{thm:optimal},
\begin{equation}
\bigl[ \omega_{\theta+\varepsilon}(x) - \omega_\theta(x) \bigr]^2 \leq 4 \left[ 1 - B(\Pb_\theta,\Pb_{\theta+\varepsilon}) \right] \left[ \omega_{\theta+\varepsilon}\left( (x-c)^2 \right) + \omega_\theta\left( (x-c)^2 \right) + 2 \tau^2 \right] .
\end{equation}
Since $\theta \mapsto \xi_\theta$ is differentiable in norm and $x$ is bounded, dividing by $\varepsilon^2$ and letting $\varepsilon \to 0$ the left-hand side converges to $\dot\omega_\theta(x)^2$, while the last bracket converges to $2 \Var_{\omega_\theta}(x) + 2 \tau^2$. This is the last inequality of~\eqref{eq:gauss-bound}. Letting $\tau \to 0$ and taking the supremum over $x$ gives $\Fs_\M(\theta)$ by Cor.~\ref{cor:equivforms}. For unbounded $L$, the stated domain assumptions give the same two limits, as in the proof of Thm.~\ref{thm:optimal}.
\end{proof}

\begin{remark}[Achieving the bound with Fewster--Verch probes]
\label{rem:FVattain}
For the free scalar field, the bound can be approached by measurements that are induced by probes in the sense of Remark~\ref{rem:FV}. We recall from Lemma~\ref{lem:fisher-bound-supremum} that $\Fs_\M(\theta)$ is the supremum of the classical Fisher information over projective measurements with finitely many outcomes, and that for such a measurement $E$ with $\omega_\theta(E(\{k\})) > 0$ for all $k$,
\begin{equation}
\label{eq:fcl-finite}
F^\mathrm{cl}(\theta;E) = \sum_k \frac{\dot\omega_\theta\bigl( E(\{k\}) \bigr)^2}{\omega_\theta\bigl( E(\{k\}) \bigr)} , \quad \dot\omega_\theta(x) = 2 \Re \ip{\dot\xi_\theta}{x \, \xi_\theta} .
\end{equation}

For Gaussian-modulated measurements of smeared fields $L = \phi(f)$, Mandrysch and Navascu\'es~\cite{MandryschNavascues2025} show that $E^{\phi(f),\tau}$ for every $\tau > 0$ is obtained as a limit of measurements which are induced by probes coupled in a bounded region in which $f$ is localized. For coherent states these measurements approach the bound (Prop.~\ref{prop:fieldopt}). Independently of the probe model, Oeckl~\cite{Oeckl2025} obtains the spectral measure of a field operator as the limit of such regularized measurements and shows that the corresponding operations are causal, in contrast to the sharp measurement itself; see also~\cite{Oeckl2025b}. In this sense the bound of Prop.~\ref{prop:fieldopt} is approached by causal measurements and attained only in the sharp limit.

Fewster~\cite{Fewster2026} has shown that every local observable of the Klein--Gordon field has an \emph{exact} measurement scheme with compact coupling region. There, the probe is a second copy of the field, and the coupling exchanges system and probe observables localized in a given bounded region $\mathcal{O}$, so that the induced observable map is the identity there. Measuring a POVM in the probe copy of the algebra of $\mathcal{O}$ therefore induces the same POVM on the system. This result has been used in~\cite{MandryschSimmonsNavascues2026} in the form that every POVM in the von Neumann algebra $\M(\mathcal{O})$ is induced. Together with Lemma~\ref{lem:fisher-bound-supremum}, it shows that $\Fs_{\M(\mathcal{O})}(\theta)$ is the supremum of the classical Fisher information over measurements in $\M(\mathcal{O})$ that are induced by Fewster--Verch probes.

Finally, Mandrysch, Simmons and Navascu\'es~\cite{MandryschSimmonsNavascues2026} proved that every POVM $E$ with finitely many outcomes is \emph{asymptotically} realizable: there are Fewster--Verch schemes with induced POVMs $E_n$ such that $\omega(E_n(\{k\})) \to \omega(E(\{k\}))$ for all $k$ and all normal states $\omega$. Since the effects are uniformly bounded, this implies $\ip{\dot\xi_\theta}{E_n(\{k\}) \xi_\theta} \to \ip{\dot\xi_\theta}{E(\{k\}) \xi_\theta}$, and by~\eqref{eq:fcl-finite} also $F^\mathrm{cl}(\theta;E_n) \to F^\mathrm{cl}(\theta;E)$. Hence asymptotic realizability is sufficient to approach the bound. Moreover, they show that the corresponding measurement schemes are of \emph{non-demolition} type: they do not disturb observables that commute with the measured ones.

What is restricted in the Fewster--Verch framework is therefore not the set of local POVMs, but the set of state updates: instruments built from local Kraus operators can be acausal (Sorkin's impossible measurements), and there are even causal channels without a Fewster--Verch realization~\cite{MandryschSimmonsNavascues2026}. Since quantum metrology uses only outcome statistics, these restrictions do not affect the achievability of the bound.
\end{remark}

\subsection{Comparison with BKM Fisher information}

Here we will show that the BKM Fisher information obtained from the Araki--Uhlmann relative entropy dominates $\Fs_\M$ (Prop.~\ref{prop:bkm}), so the corresponding Cram\'er--Rao bound is valid but in general not achieved. We denote by $S_\mathrm{rel}(\omega \Vert \varphi)$ Araki's relative entropy of normal states~\cite{Araki1975}, normalized so that $S_\mathrm{rel}(\rho \Vert \sigma) = \Tr \rho\left( \ln\rho - \ln\sigma \right)$ for density matrices. In finite dimensions, the limit $\lim_{\varepsilon \to 0} 2 S_\mathrm{rel}(\rho_\theta \Vert \rho_{\theta+\varepsilon})/\varepsilon^2$ gives the BKM Fisher information, which is at least the SLD Fisher information~\cite{Petz1996}. However, for a general von Neumann algebra this limit might not exist everywhere since relative entropy is only convex and lower semicontinuous~\cite{Araki1975}. Nevertheless, this is enough to show that an inequality still holds.

\begin{proposition}[SLD versus BKM]
\label{prop:bkm}
Let $\theta \mapsto \omega_\theta$ be a curve of faithful normal states on a von Neumann algebra $\M$ for which $\theta \mapsto \xi_\theta$ is continuously differentiable in norm, where $\xi_\theta$ is the canonical implementing vector of $\omega_\theta$. Then it holds that
\begin{equation}
\liminf_{\varepsilon \to 0} \frac{2 S_\mathrm{rel}(\omega_\theta \Vert \omega_{\theta+\varepsilon})}{\varepsilon^2} \geq \Fs_\M(\theta).
\end{equation}
In particular, whenever the BKM Fisher information $\Fb_\M(\theta) \coloneq \lim_{\varepsilon \to 0} 2 S_\mathrm{rel}(\omega_\theta \Vert \omega_{\theta+\varepsilon})/\varepsilon^2$ exists, we have $\Fb_\M(\theta) \geq \Fs_\M(\theta)$.
\end{proposition}
\begin{proof}
Consider $x = x^* \in \M$ and the von Neumann subalgebra $\mathcal{N} \subset \M$ generated by spectral projections of $x$. By the monotonicity of the relative entropy~\cite[Thm.~4.1]{Kosaki1986}, it holds that $S_\mathrm{rel}(\omega_\theta \Vert \omega_{\theta+\varepsilon}) \geq S_\mathrm{rel}\left( \omega_\theta \rvert_\mathcal{N} \Vert \omega_{\theta+\varepsilon} \rvert_\mathcal{N} \right)$. Since $\omega_\theta$ is faithful for all $\theta$, also its restriction to $\mathcal{N}$ is faithful, and since $\mathcal{N}$ is Abelian, we have the variational formula~\cite[Thm.~9, Remark after Cor.~5]{Petz1988}
\begin{equation}
S_\mathrm{rel}\left( \omega_\theta \rvert_\mathcal{N} \Vert \omega_{\theta+\varepsilon} \rvert_\mathcal{N} \right) = \sup_{h = h^* \in \mathcal{N}} \left[ \omega_\theta(h) - \ln \omega_{\theta+\varepsilon}\left( \mathe^h \right) \right] = \sup_f \left[ \omega_\theta(f(x)) - \ln \omega_{\theta+\varepsilon}\left( \mathe^{f(x)} \right) \right] ,
\end{equation}
where the second supremum is taken over all Borel functions $f \colon \mathbb{R} \to \mathbb{R}$. Choosing $f(x) = - s \varepsilon x$ with $s \in \mathbb{R}$, we thus obtain
\begin{equation}
S_\mathrm{rel}(\omega_\theta \Vert \omega_{\theta+\varepsilon}) \geq \omega_\theta(- s \varepsilon x) - \ln \omega_{\theta+\varepsilon}\left( \mathe^{- s \varepsilon x} \right) = s \varepsilon^2 \dot \omega_\theta(x) - \frac{s^2 \varepsilon^2}{2} \left[ \omega_\theta(x^2) - \omega_\theta(x)^2 \right] + o(\varepsilon^2) ,
\end{equation}
where the expansion is justified by the differentiability in norm of the implementing vectors. Choosing $x$ such that $\Var_{\omega_\theta}(x) > 0$, it follows that
\begin{equation}
\begin{split}
\liminf_{\varepsilon \to 0} \frac{2 S_\mathrm{rel}(\omega_\theta \Vert \omega_{\theta+\varepsilon})}{\varepsilon^2} &\geq \sup_{s \in \mathbb{R}} \left[ 2 s \dot \omega_\theta(x) - s^2 \left[ \omega_\theta(x^2) - \omega_\theta(x)^2 \right] \right] \\
&= \sup_{s \in \mathbb{R}} \left[ \frac{\dot \omega_\theta(x)^2}{\Var_{\omega_\theta}(x)} - \frac{\bigl[ \dot \omega_\theta(x) - s \Var_{\omega_\theta}(x) \bigr]^2}{\Var_{\omega_\theta}(x)} \right] ,
\end{split}
\end{equation}
and the supremum is attained for $s = \dot \omega_\theta(x)/\Var_{\omega_\theta}(x)$. Since $x$ is otherwise arbitrary, we may take another supremum over all $x = x^*$ with $\Var_{\omega_\theta}(x) > 0$, which by Cor.~\ref{cor:equivforms} is equal to $\Fs_\M(\theta)$.
\end{proof}

Together with Lemma~\ref{lem:fisher-bound}, Prop.~\ref{prop:bkm} shows that a Cram\'er--Rao bound built on the BKM metric is valid, and that it is reached by a measurement in $\M$ only where $\Fb_\M = \Fs_\M$. In~\cite{Chatterjee2026} the BKM form is evaluated on the reflected-difference tangent of a modularly paired family. Applied to the same curve, Thm.~\ref{thm:exist} gives the SLD value, which is smaller by Prop.~\ref{prop:bkm}, and which by Thm.~\ref{thm:optimal} is the one accessible to local measurements. For coherent states of the free field on the wedge we also show that the inequality is strict (Prop.~\ref{prop:bkm-wedge}).

\section{The free scalar field in the wedge}
\label{sec:wedge}

In this section, we turn to a simple, explicit and non-trivial example, namely a massive scalar field in a wedge, to calculate explicitly the quantum SLD Fisher information for type III von Neumann algebras. We consider a family of unitarily excited states, but with the unbounded field operator as generator, and thus Cor.~\ref{cor:unitary-unbounded} has to be used.

Since the quantization of a free scalar field is well-known (see for example~\cite{EckmannOsterwalder1973,FiglioliniGuido1989,KayWald1991}), we only give the formulas that are required to establish the notation. Let $\mathcal{H}_1$ be the single particle Hilbert space, and $\mathcal{H} = \Gamma(\mathcal{H}_1)$ the corresponding Fock space with vacuum vector $\Omega$. For $h \in \mathcal{H}_1$, let $\phi(h)$ be the field operator and $W(h) = \mathe^{\mathi \phi(h)}$ the corresponding Weyl operator, which we normalize such that $\phi(h) \Omega = h$ and $\left\langle \Omega, W(h) \Omega \right\rangle = \mathe^{- \norm{ h }^2/2}$. For $g,h \in \mathcal{H}_1$, the product of two Weyl operators is given by $W(g) W(h) = \mathe^{- \mathi \Im \ip{g}{h}} W(g+h)$. For a closed real subspace $H \subset \mathcal{H}_1$, the von Neumann algebra generated by Weyl operators $W(h)$ with $h \in H$ is given by $\left\{ W(h) \colon h \in H \right\}''$, and $\Omega$ is cyclic and separating for this algebra if and only if $H$ is a standard subspace~\cite{Araki1963}. By the Reeh--Schlieder theorem~\cite{ReehSchlieder1961}, this holds whenever $H = H_\mathcal{O}$ corresponds to functions supported in an open region $\mathcal{O}$ in Minkowski spacetime whose causal complement has nonempty interior. Concretely, we take
\begin{equation}
\mathcal{H}_1 = L^2\left( \mathbb{R}^3, \frac{1}{2 \omega_{\vec{p}}} \frac{\total^3 \vec{p}}{(2 \pi)^3} \right) , \quad \omega_{\vec{p}} \coloneq \sqrt{ \vec{p}^2 + m^2 } ,
\end{equation}
and the standard subspace
\begin{equation}
H_\mathcal{O} \coloneq \overline{ \left\{ q f \colon f \in C^\infty_\mathrm{c}\left( \mathcal{O}, \mathbb{R} \right) \right\} }^{\,\mathbb{R}} , \quad (q f)(\vec{p}) \coloneq (\mathcal{F} f)(\omega_{\vec{p}}, \vec{p}) ,
\end{equation}
where $q$ is the one-particle map, the restriction of the Fourier transform of $f$ to the mass shell $p^0 = \omega_{\vec{p}}$. The corresponding Weyl algebra is given by
\begin{equation}
\label{eq:m-o-def}
\M(\mathcal{O}) \coloneq \left\{ W(h) \colon h \in H_\mathcal{O} \right\}'' ,
\end{equation}
and it turns out that in this case, the modular operator $\Delta_\Omega$ and modular conjugation $J_\Omega$ of $( \M(\mathcal{O}), \Omega )$ are the second quantizations $\Gamma(\Delta_H)$, $\Gamma(J_H)$ of the single particle modular objects of the standard subspace $H_{\mathcal{O}}$~\cite{EckmannOsterwalder1973,FiglioliniGuido1989}.

We would like to estimate the amplitude of a classical wave of known shape. The corresponding family of states consists of coherent excitations of the Minkowski vacuum, the vector states with implementing vector $\xi_\theta = W(\theta h) \Omega$. This is analogous to the unitary family of Cor.~\ref{cor:unitary}, with the difference that the generator $\phi(h)$ is an unbounded operator. Physically, such a state is produced by a linear coupling of the quantum field $\phi$ to a classical source $j = \theta f$, where $f \in C^\infty_\mathrm{c}(\mathcal{O}, \mathbb{R})$ is a real smooth function such that $q f = h$ (see for example \cite{FullingLandulfoMatsas2020}). That is, we consider an observer with access to the region $\mathcal{O}$ (and hence the algebra $\M(\mathcal{O})$) only, estimating the unknown signal strength $\theta$ of a signal with known template $h$. We will see that being restricted to the region $\mathcal{O}$ will inevitably lead to a loss of precision of the measurement, namely to a smaller quantum Fisher information and hence to a larger minimal variance of any estimator for $\theta$.

For simplicity, we consider the standard right wedge region $\mathcal{W} = \{ x \in \mathbb{R}^{1,3} \colon x^1 > \abs{x^0} \}$, for which the modular objects are known by the Bisognano--Wichmann theorem~\cite{BisognanoWichmann1975,BisognanoWichmann1976}. Namely, let $K$ be the generator of boosts preserving the wedge, acting on $h \in H_\mathcal{W}$ as
\begin{equation}
\left( \mathe^{\mathi t K} h \right)(\vec{p}) = h\left( p^1 \cosh t - \omega_{\vec{p}} \sinh t, p^2, p^3 \right).
\end{equation}
Then we have $\ln \Delta_\mathcal{W} = - 2 \pi K$, which is an unbounded self-adjoint operator. By Lemma~\ref{lem:commutant}, the role of the ancilla is played in general by the commutant $\M'$. For the algebra $\M(\mathcal{W})$, this is the algebra $\M(\mathcal{W}')$ of the complementary left wedge region $\mathcal{W}' \coloneq \{ x \in \mathbb{R}^{1,3} \colon x^1 < - \abs{x^0} \}$, since for wedges Haag duality holds by the Bisognano--Wichmann theorem~\cite{BisognanoWichmann1975,BisognanoWichmann1976}.

\begin{theorem}[Free field]
\label{thm:free-field-wedge}
Let $\M(\mathcal{W})$ be the algebra of the wedge, and $\omega_\theta$ be the vector state on $\M(\mathcal{W})$ with implementing vector $\xi_\theta = W(\theta h) \Omega$ for some fixed $h \in H_\mathcal{W}$. Then it holds that
\begin{equation}
\Fs_{\M(\mathcal{W})} = 4 \norm{ \tanh(\pi K)\, h }^2 \leq 4 \, \norm{ h }^2 .
\end{equation}
\end{theorem}
\begin{proof}
For $h \in H_\mathcal{W}$, the field operator $\phi(h)$ is affiliated with the wedge algebra $\M(\mathcal{W})$ and the vacuum $\Omega$ is in its domain. Hence Cor.~\ref{cor:unitary-unbounded} is applicable and yields
\begin{equation}
\Fs_{\M(\mathcal{W})} = 4 \norm*{ \frac{1 - \Delta_\mathcal{W}}{1 + \Delta_\mathcal{W}} \, \phi(h) \Omega }^2 = 4 \norm{ \tanh(\pi K) \, h }^2 ,
\end{equation}
where we used that $\ln \Delta_\mathcal{W} = - 2 \pi K$ by the Bisognano--Wichmann theorem~\cite{BisognanoWichmann1975,BisognanoWichmann1976}. By spectral calculus, we have $\norm{ \tanh(\pi K) } \leq 1$.
\end{proof}

Can this bound be achieved? We consider the Gaussian-modulated measurements $E^{L,\tau}$~\eqref{eq:gauss-measurement} with $L = \phi(g)$ and a vector $g \in H_\mathcal{O}$ corresponding to an open region $\mathcal{O}$ whose causal complement has nonempty interior (which could be the wedge $\mathcal{W}$). The smeared field $\phi(g)$ is then affiliated with $\M(\mathcal{O})$, and we obtain
\begin{proposition}[Optimal field measurements]
\label{prop:fieldopt}
Let $\mathcal{O}$ be an open region of Minkowski spacetime whose causal complement has nonempty interior, and let $\M(\mathcal{O})$ be the associated von Neumann algebra of the free scalar field defined by Eq.~\eqref{eq:m-o-def}. Let $\omega_\theta$ be the vector state on $\M(\mathcal{O})$ with implementing vector $\xi_\theta = W(\theta h) \Omega$ for some fixed $h \in H_\mathcal{O}$. For the Gaussian-modulated measurements $E^{\phi(g),\tau}$ with $g \in H_\mathcal{O}$, it then holds that
\begin{equation}
F^\mathrm{cl}\left( \theta; E^{\phi(g),\tau} \right) = \frac{\left( 2 \Im \ip{h}{g} \right)^2}{\norm{ g }^2 + \tau^2} .
\end{equation}
In particular, the supremum of the classical Fisher information $F^\mathrm{cl}\left( \theta; E^{\phi(g),\tau} \right)$ over all Gaussian-modulated measurements, namely over all $\tau > 0$ and $g \in H_\mathcal{O}$, equals the quantum SLD Fisher information $\Fs_{\M(\mathcal{O})}$. Moreover, a sharp measurement of the field operator $\phi(k)$ with $k = \pm \Pi_0 \mathi h$, where $\Pi_0$ is the real-orthogonal projection on $H_\mathcal{O}$, achieves $\Fs_{\M(\mathcal{O})}$.
\end{proposition}
\begin{proof}
Since we consider a free scalar field and the coherent vector state with implementing vector $W(\theta h) \Omega$, the outcome distribution $\Pb_\theta \coloneq \omega_\theta \circ E^{\phi(g),\tau}$ is a Gaussian probability measure on $(\mathbb{R},\mathcal{B}(\mathbb{R}))$. Using the Weyl relations, its mean is given by~\eqref{eq:gauss-measurement-mean-variance}
\begin{equation}
\begin{split}
\int_\mathbb{R} x \total \Pb_\theta(x) &= \omega_\theta\left( \phi(g) \right) = \left\langle W(\theta h) \Omega, \phi(g) W(\theta h) \Omega \right\rangle \\
&= - \mathi \left[ \partial_s \left\langle W(\theta h) \Omega, W(s g) W(\theta h) \Omega \right\rangle \right]_{s = 0} \\
&= - \mathi \left[ \partial_s \left( \mathe^{2 \mathi \theta s \, \Im \ip{h}{g}} \mathe^{- s^2 \norm{ g }^2/2} \right) \right]_{s = 0} = 2 \theta \, \Im \ip{h}{g} ,
\end{split}
\end{equation}
and analogously we compute that its second moment is given by~\eqref{eq:gauss-measurement-mean-variance}
\begin{equation}
\int_\mathbb{R} x^2 \total \Pb_\theta(x) = \omega_\theta\left( \left[ \phi(g) \right]^2 \right) + \tau^2 = 4 \theta^2 \left( \Im \ip{h}{g} \right)^2 + \norm{ g }^2 + \tau^2 .
\end{equation}
For a Gaussian probability distribution, the density (with respect to Lebesgue measure on $\mathbb{R}$) is given by
\begin{equation}
p_\theta(x) = \frac{\exp\left[ - \frac{\left( x - \mu\left( \Pb_\theta \right) \right)^2}{2 \Var\left( \Pb_\theta \right)} \right]}{\sqrt{2 \pi \Var\left( \Pb_\theta \right)}} = \frac{\exp\left[ - \frac{\left( x - 2 \theta \, \Im \ip{h}{g} \right)^2}{2 (\norm{ g }^2 + \tau^2)} \right]}{\sqrt{2 \pi (\norm{ g }^2 + \tau^2)}} .
\end{equation}
Hence, for the classical Fisher information~\eqref{eq:fcl-expression}, we obtain as required
\begin{equation}
F^\mathrm{cl}\left( \theta; E^{\phi(g),\tau} \right) = 4 \norm{ \partial_\theta \sqrt{p_\theta} }^2_{L^2(\mathbb{R})} = 4 \frac{\left( \Im \ip{h}{g} \right)^2}{\norm{ g }^2 + \tau^2} .
\end{equation}

Since $\abs{ \Im \ip{h}{g} } \leq \norm{ h } \norm{ g }$, we obtain $\abs{ F^\mathrm{cl}\left( \theta; E^{\phi(g),\tau} \right) } \leq 4 \norm{ h }^2$, and the Fisher information is bounded for all $g \in H_\mathcal{O}$ and all $\tau \geq 0$. For the quantum SLD Fisher information, employing Cor.~\ref{cor:unitary-unbounded} we obtain
\begin{equation}
\Fs_{\M(\mathcal{O})} = 4 \norm{ \Pi_0 \mathi \phi(h) \Omega }^2 = 4 \norm{ \Pi_0 \mathi h }^2 ,
\end{equation}
where $\Pi_0$ is the real-orthogonal projection on $H_\mathcal{O}$. Since $g \in H_\mathcal{O}$, we compute
\begin{equation}
\Im \ip{h}{g} = \Re \ip{\mathi h}{g} = \Re \ip{\mathi h}{\Pi_0 g} = \Re \ip{\Pi_0 \mathi h}{g} ,
\end{equation}
and it follows that
\begin{equation}
\Fs_{\M(\mathcal{O})} - F^\mathrm{cl}\left( \theta; E^{\phi(g),\tau} \right) = 4 \frac{\norm{ \Pi_0 \mathi h }^2 \left( \norm{ g }^2 + \tau^2 \right) - \left( \Re \ip{\Pi_0 \mathi h}{g} \right)^2}{\norm{ g }^2 + \tau^2} \geq 0 .
\end{equation}
Clearly, equality is achieved for $\tau = 0$ and $g = \pm \Pi_0 \mathi h$, which corresponds to a sharp measurement of the field operator $\phi(g)$.
\end{proof}

How can this result be interpreted? We first note that relative to the global value $\Fs_\mathrm{global} = 4 \norm{ h }^2$ of the Fisher information, the information lost to the local observer on the wedge is given by
\begin{equation}
\Fs_\mathrm{global} - \Fs_{\M(\mathcal{W})} = 4 \norm*{\left( 1 - \Pi_\mathcal{W} \right) \mathi h }^2 = 4 \norm*{ \frac{1}{\cosh(\pi K)} h }^2 ,
\end{equation}
which is the component of the signal that can be generated from the complementary left wedge. To give further interpretation to the result, we consider the measure $\mu_h$ defined by
\begin{equation}
\mu_h(B) \coloneq \left\langle h, \left( 1 - \mathe^{-2 \pi K} \right) E^K\left( B \cap (0,\infty) \right) \, h \right\rangle
\end{equation}
for any Borel set $B \subset \mathbb{R}$. To understand its physical significance, we first note that $- 2 \pi J K J = J \ln \Delta_\mathcal{W} \, J = \ln \Delta_\mathcal{W}^{-1} = 2 \pi K$ and $h = S h$ for $h \in H_\mathcal{W}$, such that the spectral measure of $K$ applied to $h$ satisfies
\begin{equation}
\label{eq:kms-spectral}
\begin{split}
\total \norm{ E^K(k) h }^2 &= \total \norm{ J E^K(-k) J S h }^2 = \total \norm{ E^K(-k) \Delta_\mathcal{W}^{1/2} h }^2 \\
&= \total \norm{ E^K(-k) \mathe^{- \pi K} h }^2 = \mathe^{2 \pi k} \total \norm{ E^K(-k) h }^2 .
\end{split}
\end{equation}
We then compute the commutator between a field $\phi(h)$ and its boosted copy $\phi\left( \mathe^{\mathi s K} h \right)$:
\begin{equation}
\begin{split}
\left[ \phi(h), \phi\left( \mathe^{\mathi s K} h \right) \right] &= 2 \mathi \Im \left\langle h, \mathe^{\mathi s K} h \right\rangle = 2 \mathi \left\langle h, \sin(s K) h \right\rangle = 2 \mathi \int_{-\infty}^\infty \sin(s k) \total \norm{ E^K(k) h }^2 \\
&= 2 \mathi \int_0^\infty \sin(s k) \left( 1 - \mathe^{-2 \pi k} \right) \total \norm{ E^K(k) h }^2 = 2 \mathi \int_0^\infty \sin(s k) \total \mu_h(k) .
\end{split}
\end{equation}
This commutator is a multiple of the identity, namely the classical symplectic form between the wave $h$ and its boosted copy, independent of the state, and $\mu_h$ is the measure of its spectrum in boost frequency $k$. For the quantum SLD Fisher information, we employ Thm.~\ref{thm:free-field-wedge} and the relation~\eqref{eq:kms-spectral} to obtain
\begin{equation}
\label{eq:sld-wedge-spectral}
\begin{split}
\Fs_{\M(\mathcal{W})} &= 4 \norm{ \tanh(\pi K)\, h }^2 = 4 \int_{-\infty}^\infty \tanh^2(\pi k) \total \norm{ E^K(k) h }^2 \\
&= 4 \int_0^\infty \tanh^2(\pi k) \left( 1 + \mathe^{-2 \pi k} \right) \total \norm{ E^K(k) h }^2 = 4 \int_0^\infty \tanh(\pi k) \total \mu_h(k) .
\end{split}
\end{equation}
In terms of the Bose occupation number $n(k) = 1/(\mathe^{2 \pi k} - 1)$, this can also be written as
\begin{equation}
\Fs_{\M(\mathcal{W})} = 4 \int_0^\infty \frac{1}{2 n(k) + 1} \total \mu_h(k) ,
\end{equation}
and analogously we obtain
\begin{equation}
\Fs_\mathrm{global} = 4 \norm{ h }^2 = 4 \int_0^\infty \coth(\pi k) \total \mu_h(k) = 4 \int_0^\infty \left[ 2 n(k) + 1 \right] \total \mu_h(k) .
\end{equation}

For an observer in $\mathcal{W}$ with proper acceleration $a$, the boost parameter is $a \, \tau_\mathrm{p}$ with the proper time $\tau_\mathrm{p}$ of the observer. Therefore, the boost frequency $k$ corresponds to the proper frequency $\omega = a k$, and it holds that $2 \pi k = \omega/T_\mathrm{U}$ with the Unruh temperature $T_\mathrm{U} = a/(2 \pi)$~\cite{Unruh1976}. Hence, all observers observe the same spectrum after normalization to their measured temperature. The state-independent commutator carries the signal, with spectral weight $\mu_h$, and the Unruh fluctuations of the vacuum restricted to $\mathcal{W}$ act as noise, with weight $(2n+1) \, \mu_h$. The wedge observer's Fisher information is the signal divided by the noise factor $2n+1$, whereas the global observer, for whom the vacuum is pure, obtains the signal multiplied by this factor. For each boost frequency, we thus obtain the Fisher information of a displaced thermal mode at temperature $T_\mathrm{U}$: components with $\omega \ll T_\mathrm{U}$, hence $k \ll 1$ have $n(k) \gg 1$ and are thus almost invisible to the wedge observer, contributing very little to the observer's Fisher information $\Fs_{\M(\mathcal{W})}$. On the other hand, components with $\omega \gg T_\mathrm{U}$, hence $k \gg 1$ and $n(k) \ll 1$ contribute a lot to $\Fs_{\M(\mathcal{W})}$ and are fully visible. In Kubo's normalization~\cite{Kubo1957}, the dissipative susceptibility with respect to boost time is $\chi''(k) \total k = \pi \total \mu_h(k)$, such that
\begin{equation}
\Fs_{\M(\mathcal{W})} = \frac{4}{\pi} \int_0^\infty \tanh(\pi k) \, \chi''(k) \total k .
\end{equation}
This is exactly the relation between quantum Fisher information and dynamic susceptibility found by Hauke, Heyl, Tagliacozzo and Zoller for Gibbs states~\cite{HaukeHeylTagliacozzoZoller2016}, with the Gibbs state replaced by the vacuum $\Omega$, the Hamiltonian by the boost generator $K$, and the temperature by the Unruh one $T_\mathrm{U}$. Here it holds on a type III$_1$ factor, without density matrices. The physics, however, is different: a Gibbs state is mixed because of ignorance about a heat bath. The restriction $\Omega \rvert_{\M(\mathcal{W})}$ is the marginal of the pure Minkowski vacuum, mixed because of its entanglement with the complementary wedge, which follows from the Reeh--Schlieder theorem. The suppression by $\tanh(\pi k)$ is therefore an entanglement loss: the missing part $4 \norm{ (1 - \Pi_\mathcal{W}) \mathi h }^2$ is carried by the correlations between $\mathcal{W}$ and $\mathcal{W}'$.

Finally, we would like to compare the result with the BKM Fisher information. In Prop.~\ref{prop:bkm} we proved in general that the BKM Fisher information is never smaller than the SLD Fisher information. The wedge is a nice example in which both quantities are known explicitly, so that this can be seen directly. Moreover, it gives an example in which the BKM metric is strictly greater than the SLD metric, so the inequality of Prop.~\ref{prop:bkm} is in general not an equality.

For the coherent family $\xi_\theta = W(\theta h) \Omega$ with $h \in H_\mathcal{W}\cap D(\abs{K}^{1/2})$, the relative entropy is known explicitly~\cite{CasiniGrilloPontello2019,Longo2019,BostelmannCadamuroDelVecchio2022}: with our normalization $\ln \Delta_\mathcal{W} = - 2 \pi K$ it reads $S_\mathrm{rel}(\omega_\theta \Vert \omega_{\theta+\varepsilon}) = - \varepsilon^2 \ip{h}{\ln \Delta_\mathcal{W} \, h} = 2 \pi \varepsilon^2 \ip{h}{K h}$. Hence the limit in Prop.~\ref{prop:bkm} exists and is given by
\begin{equation}
\label{eq:bkm-wedge}
\Fb_{\M(\mathcal{W})} = 4 \pi \ip{h}{K h} = 4 \pi \int_{-\infty}^\infty k \total \norm{ E^K(k) h }^2 ,
\end{equation}
which is the BKM susceptibility evaluated in~\cite{Chatterjee2026}.
\begin{proposition}[SLD versus BKM on the wedge]
\label{prop:bkm-wedge}
For $h \in H_\mathcal{W} \cap D(\abs{K}^{1/2})$, it holds that
\begin{equation}
\Fb_{\M(\mathcal{W})} = 4 \pi \int_0^\infty k \total \mu_h(k) \geq \Fs_{\M(\mathcal{W})} ,
\end{equation}
and the inequality is strict if $h \neq 0$.
\end{proposition}
\begin{proof}
Using the relation~\eqref{eq:kms-spectral}, we obtain
\begin{equation}
\int_{-\infty}^\infty k \total \norm{ E^K(k) h }^2 = \int_0^\infty k (1-\mathe^{-2 \pi k}) \total \norm{ E^K(k) h }^2 = \int_0^\infty k \total \mu_h(k) ,
\end{equation}
which shows the equality. To show that $\Fb_{\M(\mathcal{W})} \geq \Fs_{\M(\mathcal{W})}$, we use the expression~\eqref{eq:sld-wedge-spectral} for $\Fs_{\M(\mathcal{W})}$, and see that it suffices to show that $g(k) \coloneq \pi k - \tanh(\pi k) \geq 0$ for $k \geq 0$. This is however immediate from $g(0) = 0$ and $g'(k) = \pi \tanh^2(\pi k) \geq 0$. 
To show strict inequality for $h \neq 0$, we note that $g(k) > 0$ for $k > 0$, such that it suffices to show that $\mu_h \neq 0$. If $\mu_h = 0$, then $E^K((0,\infty)) h = 0$ since $1 - \mathe^{-2 \pi k} > 0$ for $k > 0$, and by the relation~\eqref{eq:kms-spectral} also $E^K((-\infty,0)) h = 0$, such that $K h = 0$. However, in rapidity coordinates on the mass shell the boosts act by translations, such that $K$ has purely absolutely continuous spectrum and in particular no eigenvalue $0$. Hence $h = 0$.
\end{proof}

\acknowledgments

AM thanks Berihu Teklu for discussions related to quantum metrology. MBF has been funded by the Deutsche Forschungsgemeinschaft (DFG, German Research Foundation) --- project no. 574438490 (``Modular Hamiltonian and relative entropy in interacting quantum field theories''). We thank Jan Mandrysch for valuable comments on the manuscript, in particular for drawing our attention to recent results on the realizability of local measurements in the Fewster--Verch framework.

\end{document}